\RequirePackage{fixltx2e}
\documentclass[aip,jcp,reprint,showpacs,singlecolumn]{revtex4-1}

\usepackage{amsmath,amsfonts}
\usepackage{graphicx}
\usepackage[acronym]{glossaries}
\usepackage{dcolumn}
\usepackage{bm}
\usepackage{ifpdf}
\usepackage{natbib}
\usepackage{wasysym}
\usepackage[byname]{smartref}
\usepackage{color}
\usepackage{lipsum}
\usepackage{nicefrac}
\usepackage[breaklinks, colorlinks, linkcolor=blue, citecolor=blue, urlcolor=blue]{hyperref}
\usepackage[table]{xcolor}
\usepackage{multirow}
\usepackage{braket}
\usepackage{float}
\usepackage{comment}
\usepackage{enumerate}
\usepackage{enumerate}

\usepackage{amsmath,amssymb,graphicx}
\usepackage{epsfig}

\newcommand{\beq}{\begin{equation}}
\newcommand{\eeq}{\end{equation}}

\newcommand{\bea}{\begin{eqnarray}}
\newcommand{\eea}{\end{eqnarray}}

\begin{document}


\title{Charge transport in molecular junctions: From tunneling to hopping with the probe technique}
\author{Michael Kilgour}
\affiliation{Chemical Physics Theory Group, Department of Chemistry, University of Toronto,
80 St. George Street Toronto, Ontario, Canada M5S 3H6}

\author{Dvira Segal}
\email{dsegal@chem.utoronto.ca}
\affiliation{Chemical Physics Theory Group, Department of Chemistry, University of Toronto,
80 St. George Street Toronto, Ontario, Canada M5S 3H6}

\date{\today}
\begin{abstract}
We demonstrate that a simple phenomenological approach can be used to
simulate electronic conduction in molecular wires under thermal effects induced by the
surrounding environment. This ``Landauer-B\"uttiker's probe technique" can properly
replicate different transport mechanisms: phase coherent nonresonant tunneling, ballistic behavior, and
hopping conduction, to provide results consistent with experiments.
Specifically, our simulations with the probe method recover the following central characteristics of charge transfer in molecular wires:
(i) The electrical conductance of short wires falls off exponentially with molecular length, a manifestation of
the tunneling (superexchange) mechanism.
Hopping dynamics overtakes superexchange in long wires demonstrating an ohmic-like behavior.
(ii) In off-resonance situations, weak dephasing effects 
facilitate charge transfer.
Under large dephasing the electrical conductance is suppressed.
(iii) At high enough temperatures, $k_BT/\epsilon_B>1/25$, with $\epsilon_B$ as the molecular-barrier height,
the current is enhanced by a thermal activation (Arrhenius) factor. However, this enhancement
takes place for both coherent and incoherent electrons and it does not readily indicate the underlying mechanism.
(iv) At finite-bias, dephasing effects impede conduction in resonant situations.
We further show that memory (non-Markovian) effects can be implemented within the Landauer-B\"uttiker's probe technique
to model the interaction of electrons with a structured environment.
Finally, we examine experimental results of electron transfer in conjugated molecular wires
and show that our computational approach can reasonably reproduce reported values to provide mechanistic information.
\end{abstract}

\maketitle

\section{Introduction}
\label{Sintro}

Understanding of charge transport mechanisms in single-molecule junctions is
essential for the realization of molecular electronic devices, as well as for elucidating
many processes in chemistry, biology, and condensed phase physics.
Examples include charge transfer in DNA \cite{Barton,BartonRev}, a process which plays a crucial role
in mutagenesis and carcinogenesis,
electron transfer reactions \cite{Kuznetsov,May,nitzan}, and electron correlation effects at the nanoscale, such as in
quantum dots and molecular junctions \cite{datta,cuevas,diventra}.

Experiments probing electron transfer (ET) rates in donor-acceptor DNA molecules
\cite{DNA,Giese,danny,Bartons}, self-assembled monolayers
and single-molecule wires
\cite{Selzer1,Selzer2,Selzer3, Weiss, Randall,Exp-Frisbie,Exp-Frisbie-jacsc,Exp-Frisbie-jacs,Exp-Frisbie-cmat,Exp-Frisbie-R,Exp-Lu, Sara,Taowire,Exp-porphyrin,Borguet,Lee,Cahen,Wandlowski,Bufon}
have demonstrated the central role of two limiting transport mechanisms:
phase coherent nonresonant tunneling (superexchange) and incoherent thermally-activated hopping.
Direct tunneling, a single step transition, dominates transport in short wires,
but becomes unlikely at large molecular lengths as tunneling rates decrease exponentially with length.
In contrast, long-range electron transfer can take place when
the transport process is broken into multiple steps, with the electron (or hole),
transiently localized on molecular sites, hopping between them.
Furthermore, in molecular wires bridging two voltage-biased electrodes,  ``field emission",
(Fowler-Nordheim tunneling) behavior develops when the applied voltage bias transforms
the tunneling barrier from a rectangular form into a triangular shape \cite{Exp-Frisbie-jacs,Sara}.

In an effort to explain experimental results, and moreover, predict molecular electronic functionality,
a plethora of theoretical and computational methodologies were developed, aiming to explore the role
of environmental effects (internal molecular motion or the surrounding matrix)
on molecular conduction. These tools can be roughly grouped into two classes:
(i) microscopic-physical modelings which are valuable for small systems, and (ii) phenomenological descriptions, compromising
the completeness and exactness of the model to enable large-scale calculations.
The first type of approaches corresponds to models in which molecular vibrations, and other many-body interactions
such as electron-electron repulsion are explicitly included in the model Hamiltonian.
The dynamics and steady-state properties of the system can then
be analyzed by a variety of treatments.
A non-exhaustive list, focusing on vibrational effects in molecular conduction, includes
density operator approaches \cite{Mitra, wegeR}, Green's function tools \cite{nitzan-vibrev}
and path integral simulations \cite{ISPI,INFPIy,Rabani}.
Since the vibrational degrees of freedom and/or electron-electron interactions are explicitly incorporated in such treatments,
simulations are restricted to minimal models with a single molecular electronic state, or a pair of states.
The Anderson-Holstein model takes into account a single
electronic level and a particular vibration. Other common modelings consider only
two molecular orbitals, HOMO and LUMO, as decisive for electronic conduction \cite{nitzan-vibrev}.

\begin{figure}[htbp]
\vspace{2mm}
\hspace{-24mm}{\hbox{\epsfxsize=115mm \epsffile{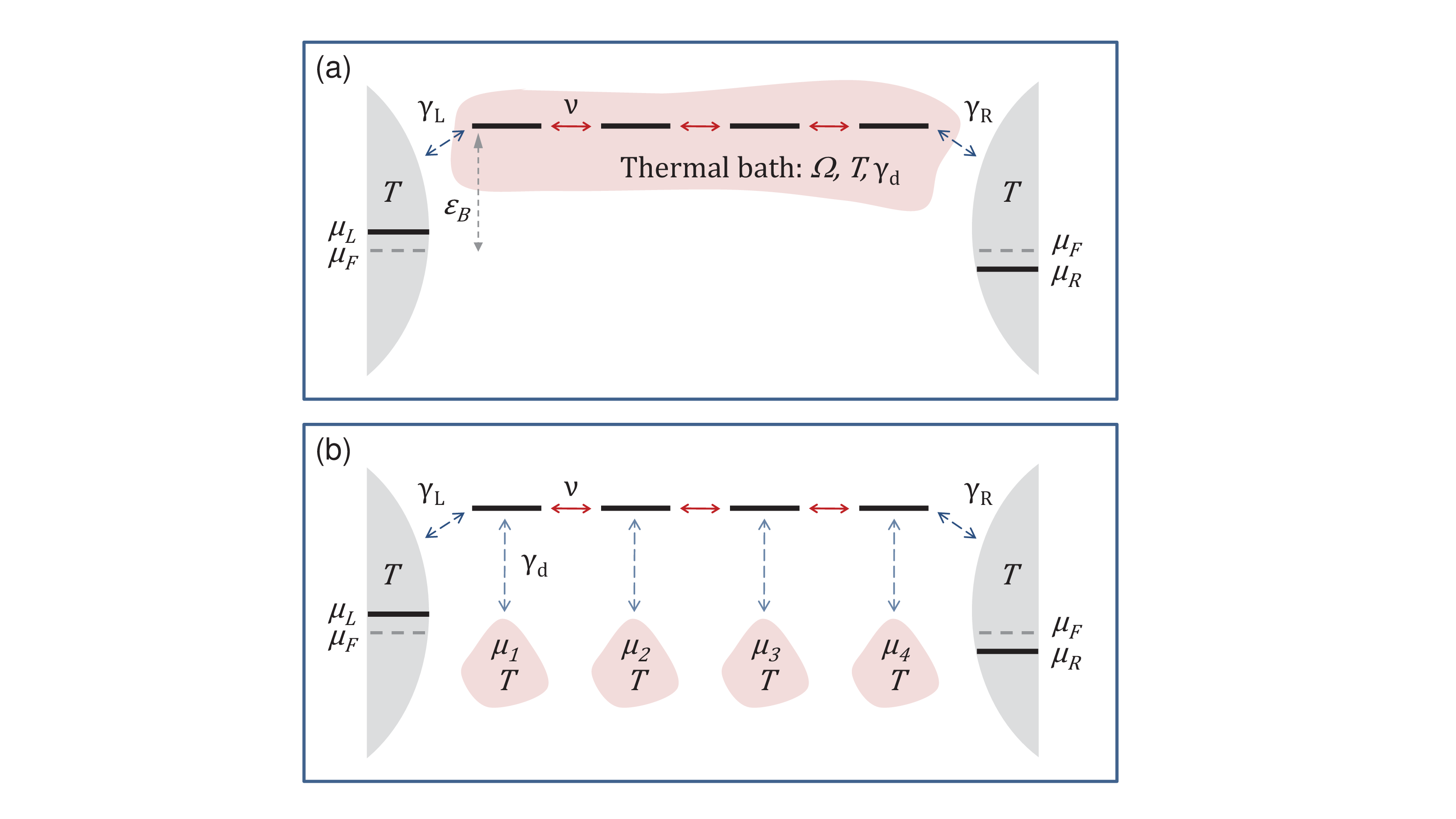}}}
\caption{ (a)
A molecular wire with $N$ electronic sites (intersite tunneling $v$ and bridge
height $\epsilon_B$) coupled to a thermal environment, represented by
the shaded region surrounding the electronic states.
The thermal bath, including molecular and external vibrational degrees of freedom,
 is characterized by its temperature $T$, spectral density function with a cutoff frequency $\Omega$
and electron-vibration coupling strength $\gamma_d$.
(b) In this work we introduce the environment
in a phenomenological manner by
using the probe technique, by attaching fictitious metal terminals to electronic sites.
The figure illustrates voltage probes, characterized by their temperature and chemical potentials which are
determined such that there is zero net charge current to each probe.
}
\label{Fig0}
\end{figure}

Complementing minimal-microscopic approaches, it is beneficial to establish
effective-phenomenological methodologies for electronic conduction with a
more favorable scaling with molecular size.
The Lindblad formalism, Redfield theory, and other kinetic equations,
are examples of such approaches \cite{Breuer, nitzan}. In these
treatments the interaction of electrons with environmental degrees of freedom
is incorporated into the dynamics via decoherence and dissipation rates which are introduced
into the equations of motion for the reduced density matrix.
Variants of such projection operator techniques have been employed extensively in the literature to model electron, proton, phonon, and exciton dynamics in condensed phases \cite{nitzan,May}.
In donor-bridge-acceptor electron transfer (DBA-ET) processes, such methods
have proven useful for describing, in a unified manner, both the tunneling and the incoherent-hopping regime
 \cite{Friesner,Mukamel,Davis, SegalET,SegalET2,Petrov1,Petrov2,Katz,Zarea, Berlin}.
More recent works rigorously examined the consistency of these effective treatments
for describing steady-state characteristics \cite{esposito06,Timm,anker13},
and applied kinetic approaches to predict electronic functionality \cite{SegalthermoE,Liang}. 

A distinct phenomenological route to implement
decoherence and inelastic effects was introduced by B\"uttiker in Ref. \cite{Buttiker-probe}.
In this ``Landauer-B\"uttiker's probe" (LBP) approach
the non-interacting Hamiltonian is augmented by probe terminals in which electrons lose their phase
memory and exchange energy with probes' degrees of freedom.
This technique, so far mostly employed in mesoscopic physics \cite{Pastawski},
e.g. to study heat to work conversion efficiency in thermoelectric devices  \cite{Dhar,Entin,saito}
and magnetotransport in quantum dot systems \cite{SalilAB,EntinB},
is particularly appealing: It allows one to model phase breaking processes e.g. in quantum dots,
while using the Landauer non-equilibrium Green's function (NEGF)
computational tool \cite{datta,cuevas,diventra, xue}. 
Beyond electronic conduction, the LBP technique has been adopted to explore
heat transport and thermal rectification in linear chains \cite{RichQ,Roy1,Roy2,LangC,Bonetto,PereiraC, Malay}
and two-dimensional constrictions \cite{Tulkki}.

The LBP method had proved itself predictive in studies of charge and heat transport in mesoscopic physics.
Can it appropriately describe environmental effects on electronic conduction in molecular junctions?
Recently, this technique has been applied
to explore the role of decoherence on quantum interference in molecular conduction  \cite{Chen-Ratner}
and to examine the connection between charge transfer kinetics and steady state currents \cite{WaldeckF},
showing that it can provide both the tunneling mechanism for short bridges,
and a ``soft"-ohmic distance dependence for longer chains.
However, a careful exploration of the LBP method in molecular electronic configurations is missing.
Our objective here is to employ the LBP technique
and examine whether it can yield results consistent with our comprehensive knowledge of environmental-assisted electron transfer
effects in molecular electronic junctions.
Particularly, the technique should depict the turnover in transport, from tunneling to hopping,
with increasing molecular size, temperature, and dephasing rates.

Another objective of this work has been to develop an approximate analytic expression for hopping conduction
in molecular junctions.
Several works have undertaken this task in a donor-bridge-acceptor
configurations to construct the {\it electron transfer rate},
a related quantity \cite{SegalET,Katz,Zarea,Petrov1,Petrov2,Beratan,Troisi}.
However, few theoretical studies had considered this problem in a metal-molecule-metal setup
 \cite{SegalET2,NitzanETC1,NitzanETC2,Lehman,Lee},
which differs fundamentally from the DBA-ET situation:
While in the latter case the initial condition places all electrons in the donor state with a given energy,
in a metal-molecule-metal experiment many electronic states in the metals contribute,
and electrons, with a thermal distribution of energies, determine the electronic conductance.

Simulations presented in this work confirm that the LBP method can properly emulate experimentally-observed
characteristics of molecular electronic conduction in wires
with a tunable ``dephasing strength" parameter, characterizing the strength of electron-environment interactions.
Based on our simulations we suggest an analytic expression which describes hopping conduction in molecular wires.
The Landauer-NEGF technique is nowadays the leading method in modeling the operation of {\it phase-coherent} molecular
electronic devices \cite{datta, cuevas,diventra,xue}. Particularly,
recent measurements of molecular thermopower (see for example Refs. \cite{ReddyS,Reddy14})
and heat dissipation in nanojunctions \cite{Reddyheat} were explained using this method.
By establishing here the appropriateness of the LBP treatment in molecular electronic conduction problems
we advocate for the natural generalization of the Landauer-NEGF technique
to incorporate environmental-assisted transport effects by including LBPs.

This paper is organized as follows. In Sec. \ref{SMM} we present the wire model and the probe technique.
Sec. \ref{Smech} summarizes transport mechanisms in molecular junctions.
In Sec. \ref{Sres} we present numerical simulations under low applied voltage.
Based on observations, we construct an analytic expression for bath-assisted molecular electronic conduction.
The behavior of the junction under large applied voltage is described in Sec. \ref{Svol}.
In Sec. \ref{Sexp} we apply our computational method and compare findings to experimental results for
transport in long conjugated molecules. 
We conclude in Sec. \ref{Ssum}.

In this work we interchangeably refer to the molecular system as a ``molecular wire", 
``molecular junction",  or, a ``molecular bridge".
Identifying the system as a ``wire" emphasizes that molecules under consideration
include repeating units, and that they are capable of transferring
charges over long distances.
The term ``junction" highlights the geometry: the molecule is placed between two metal electrodes,
distinguishing it from the DBA situation. The title ``bridge" describes the structure assumed:
We study situations in which molecular orbitals relevant for transport lie above the Fermi energies of the two electrodes.
Molecular electronic degrees of freedom are coupled to environmental coordinates (here, mimicked by probes).
The environment may correspond to internal molecular vibrations or the solvent's degrees of freedom
(we exclude the metals' electrons from this definition).
We collectively refer to these nuclear coordinates
as a ``thermal bath", ``environment", or ``surrounding", assumed to be maintained at the temperature of the metals.

\section{Model and Method}
\label{SMM}

\subsection{Model Hamiltonian}

\label{Smodel}
We consider a molecule bridging metal electrodes with the Hamiltonian
\bea
\hat H=\hat H_W+\hat H_L+\hat H_R +\hat H_T+ \hat H_P +\hat V_P.
\label{eq:H}
\eea
It comprises the molecular wire $\hat H_W$, two electrodes $\hat H_L$ and $\hat H_R$, and a coupling Hamiltonian $\hat H_T$
allowing charge transfer between the two leads and the wire.
We further introduce (in a phenomenological manner) dephasing and inelastic effects for electrons on the molecule.
This is achieved by attaching local reservoirs ($\hat H_P$) to each site in the wire (coupling Hamiltonian $\hat V_P$),
see Fig. \ref{Fig0}.
The molecular wire includes $N$ single-level sites of energy $\epsilon_n$,
\bea
\hat H_W=\sum_{n=1}^{N}\epsilon_n \hat c_n^{\dagger}\hat c_n + \sum_{n=1}^{N-1} v_{n,n+1}\hat c_n^{\dagger}\hat c_{n+1} +h.c.
\label{eq:HW}
\eea
Here, $\hat c_n^{\dagger}$ ($\hat c_n$) are fermionic creation (annihilation) operators of electrons
on each site in the wire,
the parameters $v_{n,n+1}$ are the inter-site tunneling energies.
In what follows we consider molecules made of identical building blocks, thus we
introduce the short notation $\epsilon_B=\epsilon_n$ and $v=v_{n,n+1}$.
The metal electrodes are modeled by a Fermi sea of noninteracting electrons,
\bea
\hat H_{\nu}=\sum_{k}\epsilon_{\nu,k}\hat a_{\nu,k}^{\dagger}\hat a_{\nu,k},\,\,\,\,\, \nu=L,R.
\label{eq:metal}
\eea
$\hat a_{\nu,k}^{\dagger}$ ($\hat a_{\nu,k}$) are fermionic creation (annihilation)
operators of electrons with momentum $k$ in the $\nu$ lead.
Electrons can tunnel from the $L$ ($R$) metal to site 1 ($N$),
\bea
\hat H_T=\sum_{k} g_{L,k}\hat a_{L,k}^{\dagger}\hat c_1 + \sum_{k} g_{R,k}\hat a_{R,k}^{\dagger}\hat c_N + h.c.
\label{eq:HT}
\eea
%
In the absence of the probes, this Hamiltonian dictates phase-coherent electron dynamics,
reflected e.g., by a tunneling behavior \cite{nitzan}.
We now include $N$ probes, additional metal electrodes
\bea
\hat H_P=\sum_{n=1}^N\sum_{k} \epsilon_{n,k} \hat a_{n,k}^{\dagger}\hat a_{n,k}.
\eea
The $n$th probe can exchange particles with the $n$th site of the molecular wire,
\bea
\hat V_P=\sum_{n=1}^N\sum_{k}g_{n,k}\hat a_{n,k}^{\dagger}\hat c_{n} +h.c.
\eea
Here $\hat a_{n,k}^{\dagger}$ ($\hat a_{n,k}$) are fermionic creation (annihilation) operators for an electron in the
$n=1,2,...,N$ probe with momentum $k$, $g_{n,k}$ are the
tunneling energies from the $n$th molecular site into the $n$th probe.
To eliminate charge leakage processes from the molecular wire into the probes we
enforce certain conditions on conduction.
In Sec. \ref{Sprobe} we describe in detail two such constrains, the ``dephasing probe"  and the ``voltage probe".

\subsection{B\"uttiker's Probes technique}
\label{Sprobe}

The Landauer approach provides a simple-exact description of phase coherent quantum transport \cite{datta}.
Given its simplicity, it is appealing to use it beyond the coherent limit.
Indeed, as was shown in Ref. \cite{Buttiker-probe}, one can
implement (elastic and inelastic) scattering of electrons with other degrees of freedom, possibly phonons, photons,
and other electrons, by introducing additional terminals (probes) into the model system.
The key point here is that the parameters of these terminals, essentially their local charge distributions,
should be set in a self-consistent way such that there is no net (average) particle current between the physical system
of interest and the probes.

The probe technique can be implemented under different self-consistent conditions,
allowing us to craft electron scattering processes: elastic dephasing effects
are implemented via the ``dephasing probe", while dissipative inelastic effects are introduced through the ``voltage probe". Further, dissipation-less
inelastic scattering processes can be admitted by requiring the net average fluxes of particles and  heat between
the probe terminal and the system to vanish,  termed as the ``temperature-voltage probe".
These probes can be operated in the linear response regime, as well as far from equilibrium \cite{SalilEPJ}.

Below we employ $\nu$ to identify the $L$ and $R$ (physical) metal electrodes to which the molecule is connected.
We count the probe terminals with the index $n$ and use $\alpha$ to identify
all leads, the two metal electrodes $\nu=L,R$  and the $n=1,2,..,N$ probes.

Since the model Hamiltonian does not include interactions,
its charge transfer characteristics can be described with the Landauer-B\"uttiker formalism \cite{datta}.
The total current leaving the $L$ contact is given by
\bea
I_L=\frac{e}{2\pi\hbar}\sum_{\alpha} \int_{-\infty}^{\infty}\mathcal T_{L,\alpha}(\epsilon) \left[f_L(\epsilon)-f_{\alpha}(\epsilon)\right]d\epsilon.
\label{eq:currL}
\eea
This expression should be multiplied by a factor of $2$ to account for the spin degree of freedom.
Since magnetic effects are absent, $\mathcal T_{\alpha,\alpha'}(\epsilon)=\mathcal T_{\alpha',\alpha}(\epsilon)$.
$f_{\nu}(\epsilon)=[e^{\beta(\epsilon-\mu_{\mu})}+1]^{-1}$ are the Fermi functions in the physical electrodes, given in terms of the
inverse temperature $k_BT=\beta^{-1}$ and chemical potentials $\mu_{\nu}$.
The functions $f_{n}(\epsilon)$  are to be determined from the probe condition.

The electrical conductance is
defined as the ratio of charge current to applied voltage, $\Delta V=(\mu_L-\mu_R)/e$, 
\bea
G=I_L/\Delta V.
\label{eq:conductance}
\eea
This definition does not necessarily assume low voltages.
One often further defines a linear response electrical conductance
%
demanding the applied bias to be the smallest energy scale in the system, $|e\Delta V|< 1/\beta, D, \gamma_{\nu}, \epsilon_B, v,\gamma_d$,
with $\gamma_{\nu}$ and $\gamma_d$ defined after Eq. (\ref{eq:gamma}) and $D$ as the bandwidth of the metal leads, the
largest energy scale.
In our simulations we principally work under the low-bias condition, yet for simplicity use everywhere the definition
(\ref{eq:conductance}).

In direct analogy to Eq. (\ref{eq:currL}), the net current between the $n$th probe and the system can be written as
\bea
I_{n}=\frac{e}{2\pi\hbar}\sum_{\alpha} \int_{-\infty}^{\infty} \mathcal T_{n,\alpha}(\epsilon) \left[f_n(\epsilon)-f_{\alpha}(\epsilon)\right]d\epsilon.
\label{eq:currP}
\eea
The transmission functions in Eqs. (\ref{eq:currL}) and (\ref{eq:currP}) are obtained from
the ($N\times N$) Green's function and the hybridization matrices \cite{nitzan},
\bea
\mathcal T_{\alpha,\alpha'}(\epsilon)={\rm Tr}[\bf {\Gamma}_{\alpha}(\epsilon)G^{\dagger}(\epsilon)\Gamma_{\alpha'}(\epsilon)G(\epsilon)],
\label{eq:trans}
\eea
where the trace is performed over the $N$ states of the molecule.
The Green's function is obtained from
the inverse of a tridiagonal matrix, ${\bf A= [G^{\dagger}]^{-1}}$, with the matrix elements
\bea {\bf A}_{n,n}(\epsilon)&=& \epsilon-\epsilon_{B}
+ \frac{i}{2}[\gamma_L(\epsilon)\delta_{n,1}+\gamma_R(\epsilon)\delta_{n,N}+\gamma_n(\epsilon)]
\nonumber\\
{\bf A}_{n,n\pm1}(\epsilon)&=& -v. 
\eea
The hybridization matrices have a single nonzero value as follows,
\bea
&&[ {\bf \Gamma}_{n}(\epsilon)]_{n,n}= \gamma_{n}(\epsilon),
\nonumber\\
&&[{\bf \Gamma}_{L}(\epsilon)]_{1,1}= \gamma_L(\epsilon), \,\,\,\,
[{\bf \Gamma}_{R}(\epsilon)]_{N,N}= \gamma_R(\epsilon),
\eea
with  energies
\bea
\gamma_{\alpha}(\epsilon)=2\pi\sum_{k}|g_{\alpha,k}|^2\delta(\epsilon-\epsilon_{\alpha,k}).
\label{eq:gamma}
\eea
We work in the wide-band limit, unless otherwise stated, and take
$\gamma_{\alpha}$ as energy independent parameters. Below we assume
that all sites in the wire are similarly affected by the probes, thus we use a single parameter to identify
the probe-molecule hybridization energy, $\gamma_d=\gamma_n$.
We also define the harmonic mean $\gamma=\frac{1}{2}\gamma_L\gamma_R/(\gamma_L+\gamma_R)$ as a measure for the molecule-metal coupling.

We now describe the dephasing probe and the voltage probe conditions.
In the linear response regime at low enough temperatures the two probes act similarly
on the molecule.


{\it Dephasing Probe.}
We implement local elastic dephasing effects by forcing the {\it energy-resolved} particle currents,
between every probe and the system,
to diminish.
Mathematically, we demand the integrand in Eq. (\ref{eq:currP}) to nullify.
This condition translates into a set of $N$ linear equations for the functions $f_n(\epsilon)$,
given in terms of the Fermi distributions at the $L$ and $R$ electrodes,
\bea
&&\left[\sum_{n'}\mathcal T_{n,n'}(\epsilon) + \mathcal T_{n,L}(\epsilon)+\mathcal T_{n,R}(\epsilon) \right]f_n(\epsilon)  -\sum_{n'}\mathcal T_{n,n'}(\epsilon)f_{n'}(\epsilon)
\nonumber\\
&&=\mathcal T_{n,L}(\epsilon) f_L(\epsilon)+\mathcal T_{n,R}(\epsilon) f_R(\epsilon).
\label{eq:dephG}
\eea
We solve this linear set repeatedly, for energies
$\epsilon$ extending $-D$ to $D$, with $D$ a high energy cutoff (bandwidth). The resulting probe distributions $f_{n}(\epsilon)$,
not necessarily in the form of Fermi functions, are used in Eq. (\ref{eq:currL}) to reach the net charge current.

{\it Voltage Probe.}
Local inelastic-dissipative effects can be introduced into the wire by
demanding that the net total particle current flowing between each probe and the system, Eq. (\ref{eq:currP}), vanishes,
\bea
I_n=0.
\label{eq:vol}
\eea
Out-of-equilibrium, these $N$ equations can be solved numerically, as was done in phononic models \cite{Malay}.
Analytic results for $\mu_n$ can be reached in linear response \cite{Dhar},
\bea
&&\mu_n \sum_{\alpha}\int_{-\infty}^{\infty}\left(-\frac{\partial f_{eq}}{\partial \epsilon}\right)
\mathcal T_{n,\alpha}(\epsilon) d\epsilon
\nonumber\\
&&-
\mu_{n'}\sum_{n'}\int_{-\infty}^{\infty}\left(-\frac{\partial f_{eq}}{\partial \epsilon}\right)
\mathcal T_{n,n'}(\epsilon)d\epsilon
\nonumber\\
&&= \int_{-\infty}^{\infty}d\epsilon
\left(-\frac{\partial f_{eq}}{\partial \epsilon}\right)[\mathcal T_{n,L}(\epsilon)\mu_L
+\mathcal T_{n,R}(\epsilon) \mu_R].
\label{eq:vol1}
\eea
We can further simplify Eq. (\ref{eq:vol1})
at low enough temperatures, $1/\beta<\gamma_{\nu}$. In this limit, the transmission
function is assumed to be flat
in the vicinity of the Fermi energy, in a region where the Fermi function suffers significant changes,
$\frac{\partial f_{eq}(\epsilon)}{\partial \epsilon}\sim -\delta(\epsilon-\epsilon_F)$.

At low applied bias, the dephasing and voltage similarly operate. This can be justified by noting that
in linear response and at low enough temperatures the voltage probe 
can support energy exchange processes only
within a small energy interval, $\mu_n-\mu_F$, thus it acts similarly to the dephasing probe (see Fig. {\ref{FigTD}).

What does a dephasing probe do?
While in the local-site basis
this  probe enforces the condition  of zero {\it energy resolved} flux between the every site and its local probe,
upon transforming the Hamiltonian of the wire (\ref{eq:HW}) from the site representation to the energy basis we
find that the dephasing probe can drive transitions between molecular orbitals:
We define the molecular eigenstates by the fermionic creation operator  $\hat d_m^{\dagger}$,
$\hat H_W=\sum_m E_m \hat d_m^{\dagger}\hat d_m$,
then formally expand the local-site operators,
$\hat c_n=\sum_m\lambda_{n,m}\hat d_m$.
In the energy basis the $N$ probes are coupled to the $N$ molecular orbitals,
$\hat V_P=\sum_{n,m} \lambda_{n,m}g_{n,k}\hat a_{n,k}^{\dagger}\hat d_m$  $+h.c.$, providing a mechanism for
(environmentally-mediated) transitions between molecular electronic states.
The energy resolved current to the $n$ probe is given by
\bea
I_{n,k}= i\langle \sum_m \left( g_{n,k}^*\lambda_{n,m}^*\hat d_m^{\dagger}\hat a_{n,k} - g_{n,k}\lambda_{n,m}\hat a_{n,k}^{\dagger}\hat d_m\right) \rangle
\nonumber\\
\eea
Since under the dephasing probe we demand that $I_{n,k}=0$,
this expression reveals that the $n$ probe takes electrons of energy $\epsilon_k$ e.g. from the $m$th orbital and disperses them into the other
orbitals. For example,
If we consider a single probe and two molecular orbitals, $m_1$ and $m_2$,
the dephasing probe condition translates to the conservation law $I_{m_1\to probe}(\epsilon)=I_{probe\to m_2}(\epsilon)$,
i.e., the net energy flow to the probe is zero, but it exchanges electrons between orbitals.

Below we refer to the hybridization energy $\gamma_d$ (molecule-dephasing probes)
as ``dephasing strength".
As explained above, $\gamma_d/\hbar$ does not correspond to a ``pure dephasing" process. This situation is similar to
that reached with the perturbative (system-bath weak coupling) Redfield equation \cite{SegalET2}:
Bath-assisted dynamics had been admitted to the model  by
adding dephasing operators, acting locally on different sites
of the wire. In the energy basis of the molecule these local bath operators
were responsible for bath-induced transitions between molecular orbitals
 \cite{SegalET}.
Given the probe conditions, the current leaving the $L$ contact is identical to the current
reaching the $R$ terminal, $I_L=-I_R$.

\subsection{Structured environment}

The interaction of electrons with molecular vibrations 
is explicitly described by the following model
\bea
\hat H=\hat H_W+\hat H_{L}+\hat H_R+\hat H_T+ \hat H_{ph}+\hat H_{e-ph}.
\eea
The molecular wire and the electrodes are the same as in Eqs. (\ref{eq:HW})-(\ref{eq:HT}). The bosonic (phonon)
environment and its coupling to electrons is given by
\bea
\hat H_{ph}&=&\sum_{q,n}\omega_{q,n} \hat b_{q,n}^{\dagger}\hat b_{q,n}, \,\,\,\,
\nonumber\\
\hat H_{e-ph}&=&\sum_{q,n} h_{q,n} (\hat b_{q,n}^{\dagger}+\hat b_{q,n})\hat c_n^{\dagger}\hat c_n.
\nonumber
\eea
Here $\hat b_{q,n}^{\dagger}$ ($\hat b_{q,n}$) stands for a bosonic creation (annihilation) operator
responsible for exciting a mode of frequency $q$, coupled to the electron number operator $\hat c_n^{\dagger}\hat c_n$.
The environment and its coupling energies $h_{q,n}$ provide
the  ``spectral density" function $J_n(\omega)=\sum_{q}|h_{q,n}|^2\delta(\omega-\omega_{q,n})$, 
possibly comprising a broad (low frequency)
contribution, as well as discrete high frequency vibrational modes.
It was illustrated in many studies that structured  environments support
involved dynamics in the subsystem; essentially, the back action of the bath on the system drives ``non-Markovian"
dynamics \cite{Breuer}.

To model such effects within the LBP technique
we suggest to adopt energy-dependent probe relaxation rates, defined in Eq. (\ref{eq:gamma}).
Specifically, we use below the Debye-Drude function
\bea
\gamma_d(\epsilon)= 2\tilde \gamma_d\frac{\Omega|\epsilon|}{\epsilon^2+\Omega^2}
\label{eq:SE}
\eea
which reaches the maximum value $\tilde \gamma_d$ at the cutoff frequency, $\gamma_d(\epsilon=\Omega)=\tilde\gamma_d$.
Electrons arriving at the $n$th electronic site with energy $|\epsilon|\lesssim\Omega$ are likely to exchange
energy with the probe's degrees of freedom.
In contrast, electrons will not scatter to the probe terminals if $|\epsilon|\gg \Omega$,
and they will coherently cross the junction.
This behavior corresponds to the expected effects of electron-phonon interactions. In Sec. \ref{Results-nonmark}
 we examine the role of a finite $\Omega$ on the conductance.

\begin{figure*}[t]
 \centering
\vspace{2mm}
\hspace{0.3mm}
{\hbox{\epsfxsize=67mm \epsffile{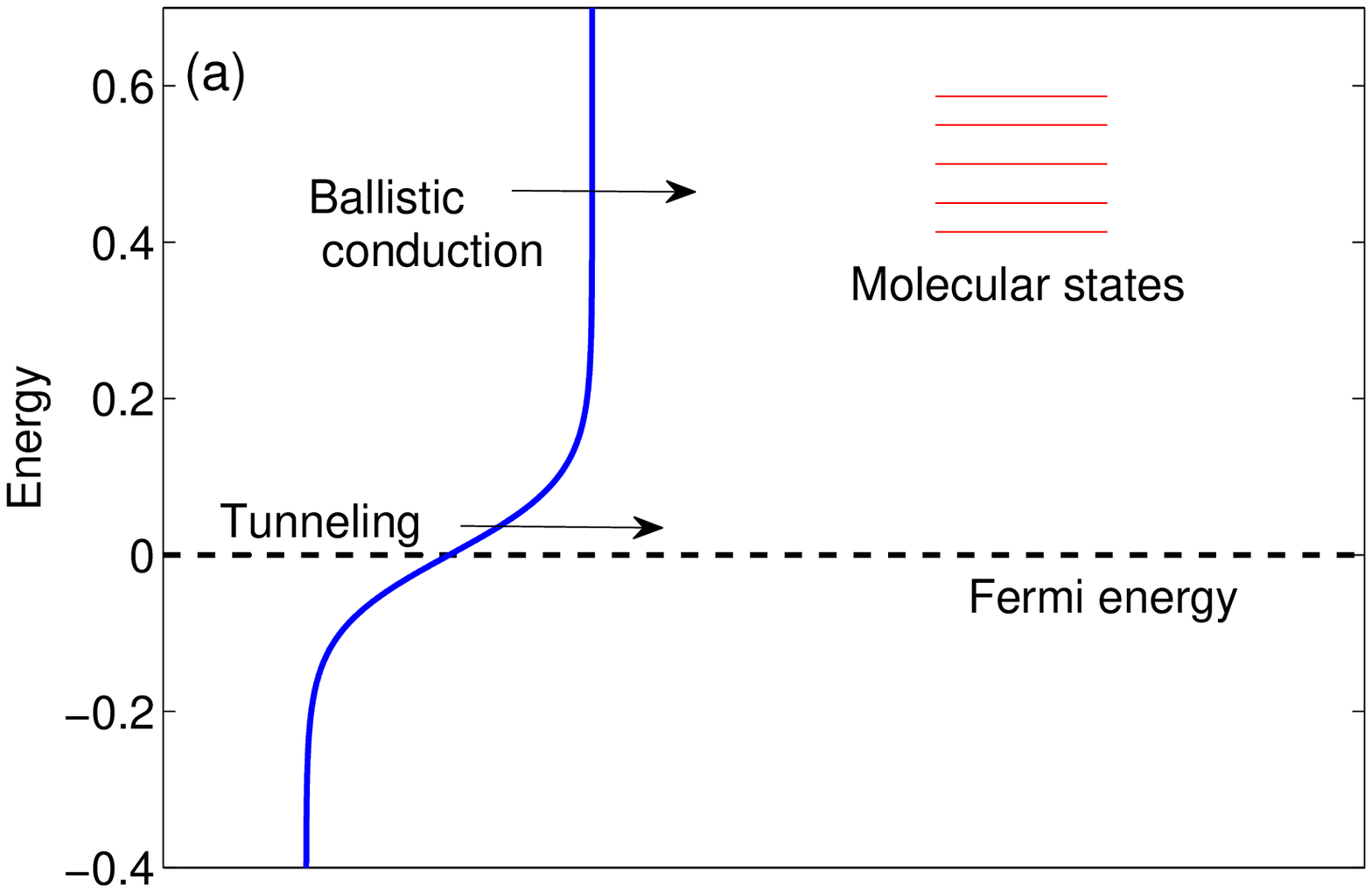}
\hspace{1mm}
\epsfxsize=72mm \epsffile{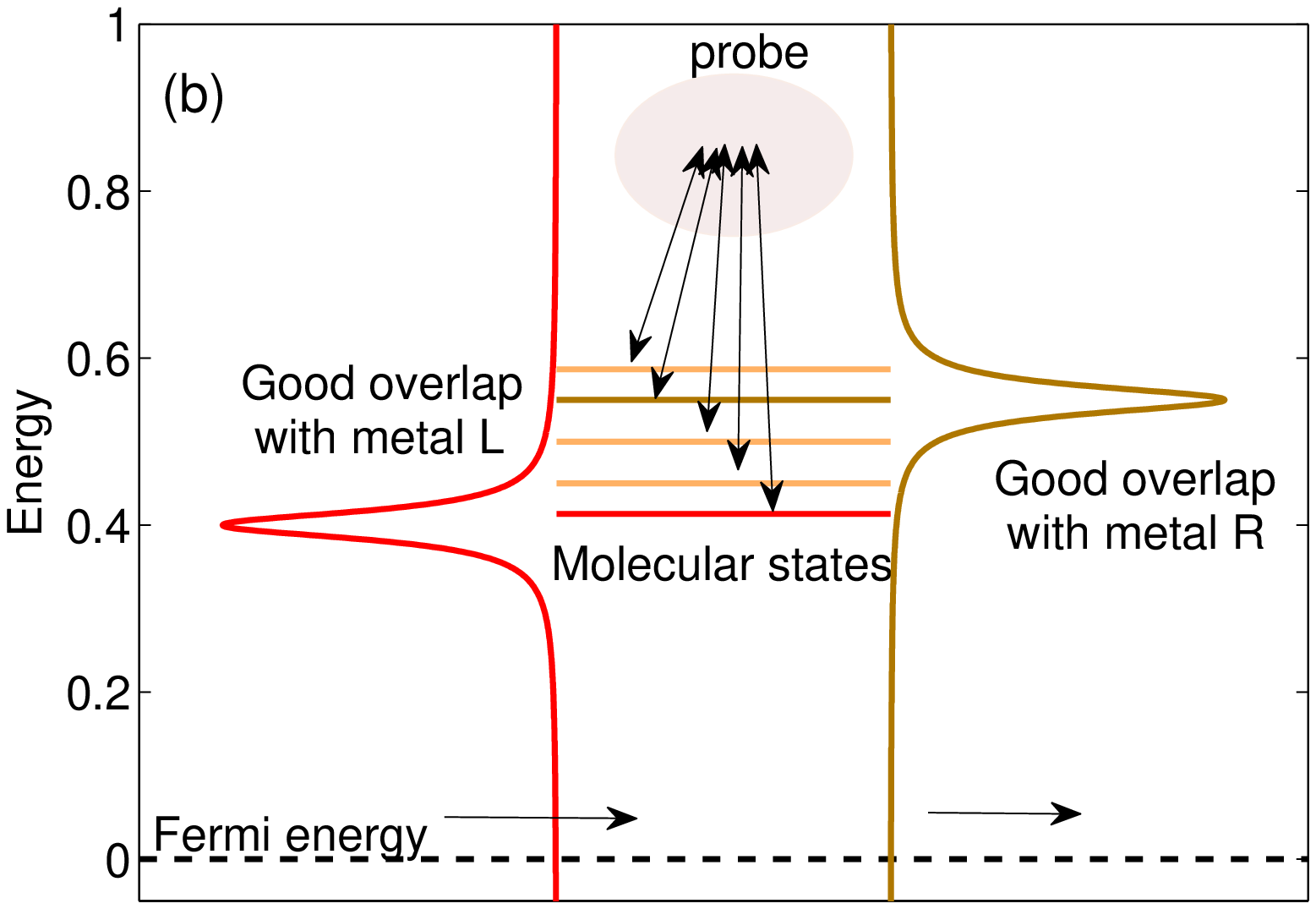}}}
\vspace{2mm}
\caption{Dominant transport mechanisms in the LBP molecular wire model.
(a) Off-resonant tunneling and (resonant) ballistic conduction when $\gamma_d=0$. The blue line
at the left end describes the Fermi distribution function of incoming electrons.
(b) Illustration of bath (probe) assisted transport, depicting for simplicity a single probe.
While all molecular orbitals are broadened, we explicitly demonstrate the broadening of two orbitals,
those that are strongly coupled to the left and right metals.
Incoming electrons of energy $\epsilon$, possibly at the tail of the broadening function, are scattered between the probes and the molecular states,
then emitted to the $R$ lead. Energy conservation is imposed by the dephasing probe.
}
\label{FigM}
\end{figure*}

\section{Transport mechanisms}
\label{Smech}

In this Section we summarize results for the electrical
conductance of molecular wires in (i) the coherent limit
when $\gamma_d=0$, see  Fig. \ref{FigM}(a), and (ii) assisted by the environment with $\gamma_d\neq0$, see
Fig. \ref{FigM}(b).

\subsection{Coherent electrons: Ballistic motion and Tunneling}
In the absence of the probes, $\gamma_d=0$, the current obeys the standard Landauer expression,
\bea
I_L=\frac{e}{2\pi\hbar}\int_{-\infty}^{\infty} d\epsilon \mathcal T_{L,R}(\epsilon)[ f_L(\epsilon)-f_R(\epsilon)].
\label{eq:Land}
\eea
Pronounced temperature dependence of transport characteristics is typically attributed to the hopping mechanism,
but the current can also strongly depend on temperature in coherent tunneling due to the
Fermi functions in this expression.
We illustrate this next.
We consider the contribution of a single-dominant narrow resonance,
\bea
\mathcal T_{L,R}(\epsilon)\sim \gamma A(\epsilon_B) \delta (\epsilon-\epsilon_B).
\label{eq:A}
\eea
Here $A$ as a dimensionless prefactor. We highlight its dependence on the bridge energetics, but
it may further depend on other molecular parameters, inter-site tunneling $v$, hybridization to the electrodes
$\gamma_{\nu}$ and molecular length $N$.
The prefactor $\gamma$, the harmonic average of $\gamma_L$ and $\gamma_R$, has been included here as the
resonance width.
Using this narrow-resonance form, Eq. (\ref{eq:Land}) reduces to
\bea
I_L&=&
\frac{e}{2\pi\hbar}\gamma A(\epsilon_B) [f_L(\epsilon_B)-f_R(\epsilon_B)]
\nonumber\\
&=&\frac{e}{2\pi\hbar}\gamma A(\epsilon_B) e^{\beta \epsilon_B}[e^{-\beta \mu_R} -e^{-\beta \mu_L}] f_L(\epsilon_B)
f_R(\epsilon_B).
\nonumber\\
\label{eq:res}
\eea
In the linear response limit, $\beta \Delta\mu<1$, we expand the functions to the lowest nontrivial order in voltage,
\bea
G&=& G_0\gamma\beta A(\epsilon_B) \frac{e^{\beta\epsilon_B}}{(e^{\beta\epsilon_B}+1)^2}
\nonumber\\
&\xrightarrow{\beta\epsilon_B\gg1}& G_0\beta\gamma A(\epsilon_B)  e^{-\beta\epsilon_B}.
\label{eq:Gr1}
\eea
The combination $\beta e^{-\beta \epsilon_B}$ suppresses the tunneling conductance at both low and high temperatures,
providing a temperature dependence distinctively different from standard Arrhenius behavior.
Here $G_0=e^2/2 \pi\hbar$; we left out the factor of $\times 2$ coming from the spin degree of freedom.

Going back to  Eq. (\ref{eq:res}), we now consider a
finite bias situation, $\beta \Delta\mu>1$.
In this case we maintain
only the larger factor $e^{\beta \Delta \mu}$ in the numerator of Eq. (\ref{eq:res}),
for $\Delta\mu>0$,
and again take the limit $\beta \epsilon_B\gg 1$. We now reach
an Arrhenius activation form for the conductance \cite{Exp-porphyrin},
\bea
G= G_0\gamma A(\epsilon_B) e^{-\beta(\epsilon_B-|e\Delta V|/2)}.
\label{eq:Gr2}
\eea
Eqs. (\ref{eq:Gr1})-(\ref{eq:Gr2})
describe contribution to conductance from thermally excited electrons, populating energies in resonance with
a narrow molecular orbital of energy $\epsilon_B$.
These electrons cross the molecule in a ballistic manner as dephasing and inelastic
terms are missing from these formulae.

Next, we broaden the resonance (\ref{eq:A}).
This allows electrons at the Fermi energy to cross the bridge, off-resonance.
We assume (i) that the applied bias is small to expand
$f_{\nu}(\epsilon)\sim f_{eq}(\epsilon)-\frac{\partial f_{eq}}{\partial \epsilon}(\mu_{\nu}-\epsilon_F)$,
and (ii) that the transmission function is almost flat (constant) close to the Fermi energy.
We then reach the following, well-known, result
\bea
I_L&=&\Delta \mu\frac{e}{2\pi\hbar} \mathcal T_{L,R}(\epsilon_F) \int_{-\infty}^{\infty} \frac{-\partial f_{eq}(\epsilon)}{\partial \epsilon} d\epsilon
\nonumber\\
&=&  \Delta \mu\frac{e}{2\pi\hbar} \mathcal T_{L,R}(\epsilon_F).
\label{eq:Ic}
\eea
The conductance is proportional to the transmission evaluated at the Fermi energy, and it does not depend on temperature,
\bea
G=G_0\mathcal T_{L,R}(\epsilon_F).
\label{eq:Gc}
\eea
In the ``deep tunneling"  regime when the barrier height is large, $\epsilon_B>v,\gamma_{\nu}$,
the transmission function can be simplified into the ``superexchange" expression \cite{nitzan}
\bea
&&\mathcal T_{L,R}(\epsilon_F) \approx
\nonumber\\
&&
\frac{\gamma_L \gamma_Rv^2}{[(\epsilon_B-\epsilon_F)^2+(\gamma_L/2)^2] [(\epsilon_B-\epsilon_F)^2+(\gamma_R/2)^2]}
\nonumber\\
&&\times
\left(\frac{v}{\epsilon_B-\epsilon_F}\right)^{2(N-2)}
\nonumber\\
&&\xrightarrow{\epsilon_B\gg\gamma_{\nu}}
\frac{\gamma_L \gamma_R}{v^2}\left(\frac{v}{\epsilon_B-\epsilon_F}\right)^{2N}
\nonumber\\
&&=  \left(\frac{\gamma_L \gamma_R}{v^2}\right)
e^{-aN \left(\frac{2}{a}\ln\big|\frac{\epsilon_B-\epsilon_F}{v}\big|\right)}.
\label{eq:exp}
\eea
Recall that we set energies relative to the Fermi energy ($\epsilon_F=0$).
Eq. (\ref{eq:exp}) indicates that the conductance decays exponentially with
molecular size, $L\equiv aN$, with $a$ as a unit length. Identifying
the tunneling decay constant by
$\kappa\equiv \frac{2}{a}\ln|\frac{\epsilon_B}{v}|$  (often denoted by $\beta$ in the literature, here
reserved for the inverse temperature),
we arrive at the familiar form
\bea
G\sim G_0e^{-\kappa L}.
\eea
Fig. \ref{FigM}(a) depicts the  ballistic [Eqs. (\ref{eq:Gr1})-(\ref{eq:Gr2})] and tunneling  [Eq. (\ref{eq:exp})]
contributions to the current.

\subsection{Bath-assisted transport}

Calculations based on perturbative master equation formalisms suggest that
long-range electrical conduction may be captured by a rational function  \cite{SegalET2},
corresponding to experimental observations of
long-range  ohmic-like conduction in molecular wires \cite{DNA,Giese, danny,Selzer1,Selzer2,Selzer3,
Exp-Frisbie,Exp-Frisbie-jacsc,Exp-Frisbie-jacs,Exp-Frisbie-cmat,Exp-Frisbie-R}. 
In Sec. \ref{Sres} we present detailed simulations using the LBP technique, and based
on observations, we construct the following form for the hopping (H) conduction,
\bea
G_{H}&\sim& G_0
A(T)
\frac{v^2}{\epsilon_B^4} \frac{\gamma_d^2}{ N + l }.
\label{eq:TH}
\eea
It is valid when the bridge is high $\epsilon_B\gg v,\gamma_d, \Delta\mu$, as well as $\epsilon_B>\gamma$, and in the range
$15<\beta\epsilon_B<30$. This corresponds to  $\epsilon_B=0.5$ eV and temperatures in the range $\pm 150$ K around room temperature.
As before, $N$ stands for the number of repeating units in the wire, $\gamma_d$ is the dephasing strength, $v$ the intersite tunneling, $\epsilon_B$ the bridge energy relative to the Fermi energy, $A(T)$ a dimensionless prefactor
which depends on temperature, possibly in a form weaker than the ``bare" Arrhenius factor, e.g.,
$A(T)\sim \beta e^{-\beta E_A}$, with an activation energy $E_A$, linearly proportional to $\epsilon_B$.
The parameter $l$ is introduced to accommodate the tunneling-to-hopping transition point,
see for example Fig. \ref{Fig2}.

Similarly to environmentally assisted DBA-ET rates, Eq. (\ref{eq:TH}) describes an ohmic conduction,
it depends on $v^2$, and as expected, grows with $\gamma_d$. However,
an exact correspondence to ET rates is missing \cite{Davis,SegalET,Petrov1,Petrov2,Beratan,Troisi,Katz,Zarea}.
While these two quantities, DBA-ET rate and the electrical conductance (examined here) are obviously related, their properties are fundamentally different in some ways:
DNA-ET processes involve electron localization on the donor and acceptor sites
 whereas in a molecular wire setup electrons are transferred through the bridge between metal electrodes.
Thus, the hybridization energy of the bridge to the metals and the thermal occupation factor
of electrons in the bulk largely determine the electrical conductance.
Furthermore, in DBA situations
the Franck-Condon factor, taking into account the energy difference between donor and acceptor states
and the solvent-induced reorganization energy, dictates the transmission rate,
while it is missing from molecular electronic conduction  \cite{NitzanETC1,NitzanETC2}.
Theoretical studies have predicted a linear relationship between DBA-ET
rates and the conductance (for molecules with high barriers), but experiments
indicate that these two quantities are not simply linearly correlated \cite{WaldeckE}, possibly
due to differences in bath-induced decoherence rates and bridge energetics \cite{WaldeckF}.

Fig. \ref{FigM}(b) schematically depicts the probe-assisted hopping conduction, Eq. (\ref{eq:TH}). While in DBA-ET situations
electrons can populate the bridge only if thermally excited, in the junction setup the molecular orbitals are broadened due to their hybridization to the metals [Lorentzian lineshapes in Fig \ref{FigM}(b)], and this broadening allows electron to occupy molecular states without
thermal activation.

We further argue that at low temperatures, equation (\ref{eq:TH}) should be generalized to
\bea
G_D=G_0\sum_{m=2,3..}^{N}\left(\frac{v^{m-1}}{\epsilon_B^m}\right)^2 F_m(\gamma_L,\gamma_R, \gamma_d,T).
\label{eq:Gdeph}
\eea
interpolating between the coherent
result (\ref{eq:exp}) using $F_N=\gamma_L\gamma_R$ and $F_{m<N}=0$
and the hopping limit (\ref{eq:TH}) when $F_2\sim A(T)\gamma_d^2/(N+l)\gg F_{m>2}$.
%


\begin{figure}[htbp]
\vspace{0mm} \hspace{0mm}
{\hbox{\epsfxsize=70mm \epsffile{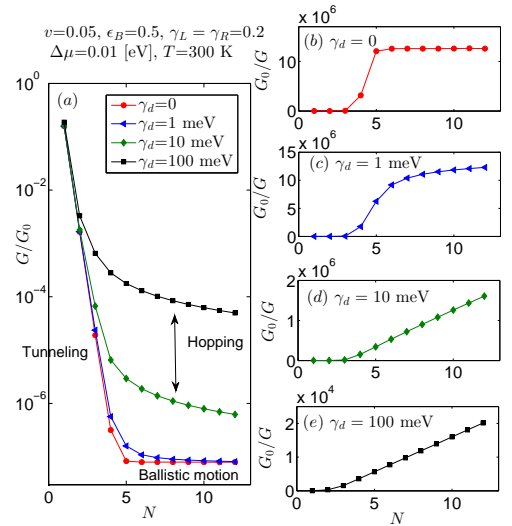}}}
\caption{Electrical conductance as a function of molecular length $N$ at $T=300$ K
using different dephasing strengths, $\gamma_d$=0, 1, 10 and 100 meV, as indicated in the figure caption,
(dephasing probe condition).
(a) A semi-logarithmic plot demonstrates that $G$ decays exponentially with length
in short chains at weak dephasing.
(b)-(e)
An inverse algebraic behavior (\ref{eq:TH}), $G^{-1}\propto N$,  is established
for long chains at large enough dephasing. 
Dominating transport mechanisms are marked in panel (a).
}
\label{Fig2}
\end{figure}


\begin{figure}[htbp]
\vspace{0mm} \hspace{0mm}
{\hbox{\epsfxsize=90mm \epsffile{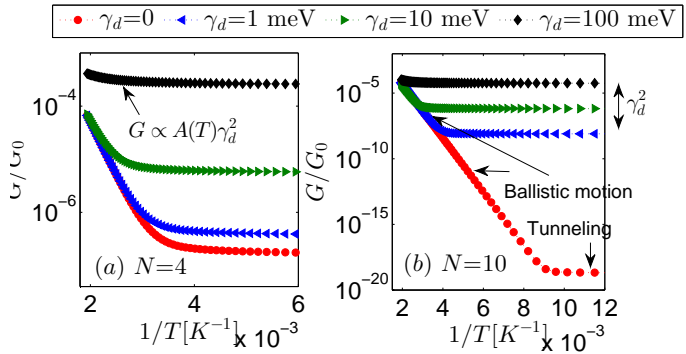}}}
\caption{
Temperature dependence of the electrical conductance for (a) $N=4$-site wire
and (b) $N=10$-site wire, demonstrating tunneling behavior at low temperatures, ballistic-resonant
activated conduction at high temperatures, and
hopping behavior when $\gamma_d\neq0$, at low-intermediate temperatures.
Parameters are $\epsilon_B=0.5$, $v=0.05$, and $\gamma_L=\gamma_R=0.2$, $\Delta \mu=0.001$ eV, dephasing probe condition.
}
\label{Fig3}
\end{figure}

\begin{figure}[htbp]
\vspace{0mm} \hspace{0mm}
{\hbox{\epsfxsize=60mm \epsffile{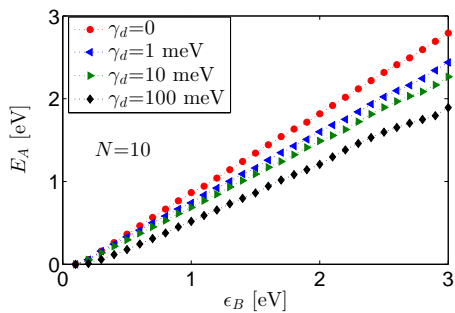}}}
\caption{
Activation energy $E_A$, as resolved from Arrhenius plots as in Fig. \ref{Fig3},
plotted against the bridge energy $\epsilon_B$,
demonstrating a linear relationship.
Parameters are indicated in Fig. \ref{Fig3}.
}
\label{Fig3b}
\end{figure}

\section{Results: tunneling to hopping conduction}
\label{Sres}

We describe here the LBP numerical simulations of molecular electronic conduction (mostly under the dephasing probe),
with the goal to construct an approximate closed-form
expression for the hopping conduction, as reported in Eq. (\ref{eq:TH}).
We focus on a uniform-symmetric system, $\epsilon_B=\epsilon_n=0.5-1$ eV, $v_{n,n\pm1}=v=0.05-0.3$ eV
and $\gamma_{L}=\gamma_R=0.2-1$ eV.
The temperature of the $L$ and $R$ leads, as well as the probes, is taken within the range $T=150-400$ K,
fixed across the junction.
In this section we restrict simulations to the low voltage regime,  $\Delta \mu=10^{-4}-0.01$ eV. The voltage
shifts the position of the Fermi energy at the contacts, $\mu_L=\mu_F+e\Delta V/2$,
$\mu_R=\mu_F-e\Delta V/2$, but bridge energies are intact.
In Sec. \ref{Svol} we study the role of a large applied voltage $\Delta \mu\sim 0.1-2$ eV
on the transport behavior.
The dephasing strength $\gamma_d$ (probe hybridization) is taken uniform along the chain.
We provide it in units of eV; when divided by $\hbar$ it translates to a rate constant.

The net current flowing under an applied bias $\Delta \mu$ is calculated in three steps:
(i) We construct the matrices ${\bf G}$ and ${\bf \Gamma_{\alpha}}$,
then the transmission coefficient between every two terminals
$\mathcal T_{\alpha,\alpha'}(\epsilon)$ from Eq. (\ref{eq:trans}).
(ii) We include environmental effects e.g. with dephasing probes.  This is achieved by
calculating (numerically) the functions $f_n(\epsilon)$ using the probe condition (\ref{eq:dephG}).
(iii) We evaluate the net-total current from the $L$ electrode via Eq. (\ref{eq:currL}).
As mentioned above, the probe conditions ensure that the current leaving the $L$ metal is identical to the current
arriving at the right electrode, $I\equiv I_L=-I_R$. For dephasing probe, moreover,
$I(\epsilon)\equiv I_L(\epsilon)=-I_R(\epsilon)$.

\subsection {Length dependence}  
\label{Results-length}

The behavior of the conductance with increasing molecular length is displayed in Fig. \ref{Fig2}
under a range of dephasing energies, $\gamma_d=0-100$ meV.
In the absence of dephasing the conductance decays exponentially with molecular length for $N\sim 1-5$, in agreement with Eq.
(\ref{eq:exp}), but beyond that it is fixed (and small), independent of size. This residual component
emerges due to incoming electrons of energy $\epsilon_B$ (occupying the tail of the Fermi function),
which are crossing the junction ballistically.
In contrast, at finite dephasing strengths and beyond $N\sim 5$
the tunneling behavior is overpowered
by a hopping-ohmic contribution $G^{-1} \propto N$.
Fig. \ref{Fig2} establishes two central results for hopping conduction:
at finite dephasing strength long-range electron transfer
follows an ohmic form, and that $G\propto \gamma_d^2$ in this region.


\subsection{Thermal activation}
\label{Results-T}

The onset of an Arrhenius behavior at high enough temperatures
is considered a central characteristic of hopping dynamics in donor-bridge-acceptor
electron transfer processes.
However, Eq. (\ref{eq:Gr2}) reveals that in metal-molecule-metal geometries
a thermally activated conduction can take place even in the coherent regime, supporting ballistic-resonant behavior.
In fact, electrons which are thermally activated in the metal are more likely to cross the junction ballistically rather than in an ohmic
fashion (after being scattered between probes). Indeed we find that a clear Arrhenius dependency in
our simulations corresponds to {\it ballistic} conduction, while a weaker
temperature dependency may correspond to thermalized-yet-off-resonant electrons, hopping across the bridge.

Figure \ref{Fig3} displays this behavior.
First, focus on the case of $\gamma_d=0$ (red line).
At very low temperatures the conductance does not depend on temperature at all
(tunneling contribution). At high temperatures, an Arrhenius behavior is observed,
with an activation energy $E_A$, indicating ballistic conduction at zero dephasing.
Fig. \ref{Fig3b} demonstrates that the activation energy
$E_A$, the slope in the Arrhenius regime, $\log G\propto -\beta$, linearly follows the bridge height $\epsilon_B$.

At finite dephasing strength the situation is rather involved.
Let us first focus on the low temperature regime of Fig. \ref{Fig3}. 
While the temperature
dependence is rather weak, it is clear that $G$ is increasing as $G\propto \gamma_d^2$.
This indicates on an activation-less dephasing assisted conductance, see Eq. (\ref{eq:TH}).
%
At high temperatures and for long wires (see for example $N=10$,
$T^{-1}=0.03$ K$^{-1}$) the conductance shows an activated form, but it does
not depend on $\gamma_d$, for $\gamma_d=1-10$ meV. This again points to a ballistic motion;
the dominant role of temperature is to enhance the population of electrons in resonance with the bridge,
with the ballistic component dominating hopping contribution.
Thermally-activated hopping conduction (\ref{eq:TH}) further shows up, but only when the dephasing strength is
high and the bridge is short, see $N=4$, $T^{-1}=0.02$ K$^{-1}$ and $\gamma_d=100$ meV.

To summarize, unlike DBA-ET rates, activated conductance in junctions does not
 expose transport mechanisms:
At low temperatures, a seemingly activation-less conductance
may reflect either deep-tunneling or dephasing-assisted conduction of deep electrons.
Similarly, activated transport at high temperatures may correspond to both ballistic 
and ohmic components.

\subsection{Dephasing}
\label{Results-deph}

The role of dephasing strength $\gamma_d$ on the conductance is explored in Fig. \ref{Fig4}, and
we observe a characteristic ``Kramers-like" turnover behavior \cite{Kramer}.
At low dephasing the conductance increases with $\gamma_d$ as it opens up a new route for electrons
to cross the bridge, by hopping between sites.
At large dephasing strengths
$\gamma_d/\epsilon_B\gtrsim 2$ the conductance drops
approximately as $\gamma_d^{-1}$. In a classical language one attributes this decay to the impediment of electrons
at strong friction, resulting in an overdamped dynamics.
This Kramers-like turnover behavior was demonstrated in other studies of electron transfer \cite{Davis,SegalET,Katz,Zarea}.
Investigations based on Redfield \cite{SegalET} and Lindblad equations
\cite{Zarea} even provided a very similar value for the turnover point.
Related trends were also observed in
recent explorations  of environmental-assisted quantum energy transport in light-harvesting biomolecules \cite{ENAQT2}.

Note that in the Kramers' theory the barrier crossing rate first increases {\it linearly} with friction strength,
then decreases as the inverse of friction. Here, we observe a more intricate behavior:
The LBP conductance increases linearly with dephasing
in short chains,  $G\propto \gamma_d$, but in long chains the enhancement follows $G\propto \gamma_d^2$, in the weak-dephasing regime.

The quadratic dependence $G\propto \gamma_d^2$ is a direct result of
the phenomenological-probe modelling of bath-assisted electron scatterings.
Unlike a genuine electron-phonon scattering process,
in which an electron is scattered {\it between} different electronic states assisted by a phonon,
the probes {\it absorb} an electron and emit it
 (possibly) at a different molecular state, to be re-absorbed by another probe.
This probe-molecule-probe transfer process of electrons
results in the $\gamma_d^2$ functional form,
 in analogy with the $\gamma_L\times \gamma_R$ term in Eq. (\ref{eq:exp}).
In short chains, the $L$ and $R$ metals-to-probes scattering is significant,
more than multiple probe-molecule-probe electron scattering, to construct the prefactor
$\gamma\gamma_d$.
While the quadratic dependence is not necessarily physical, it is
yet encouraging to find that the Kramers-like turnover is well reproduced here, in a seemingly correct value.


\begin{figure}[htbp]
\vspace{0mm} \hspace{0mm}
{\hbox{\epsfxsize=70mm \epsffile{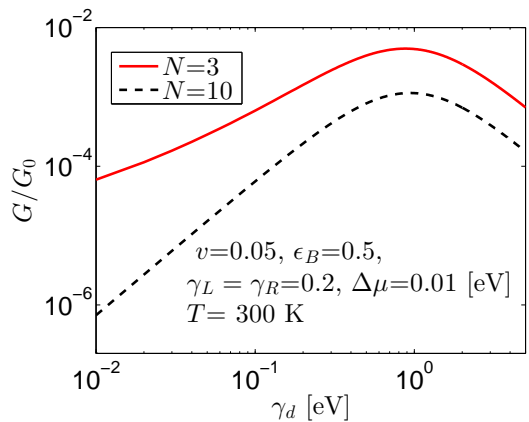}}}
\caption{
Kramers-like behavior of the electrical conductance with dephasing strength (dephasing probe).
Parameters are indicated within the figure.
}
\label{Fig4}
\end{figure}

\begin{figure}[hb]
\vspace{0mm} \hspace{0mm}
{\hbox{\epsfxsize=70mm \epsffile{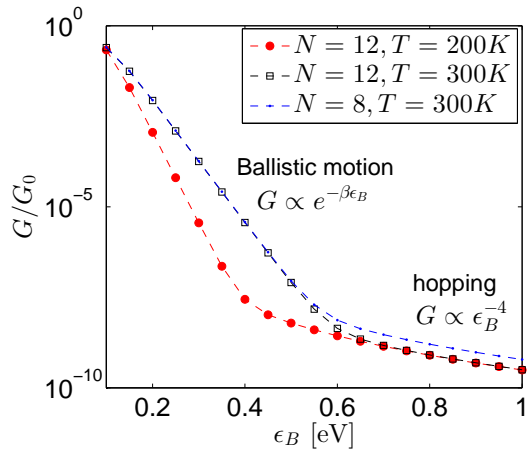}}}
\caption{
Conductance as a function of the bridge energy $\epsilon_B$.
$v=0.05$, $\gamma_{L,R}=0.2$,  $\gamma_d=10^{-3}$ (dephasing probe) and
$\Delta \mu=0.01$ eV.
}
\label{Fig5}
\end{figure}

\begin{figure*}[t]
\vspace{0mm} \hspace{0mm}
{\hbox{\epsfxsize=150mm \epsffile{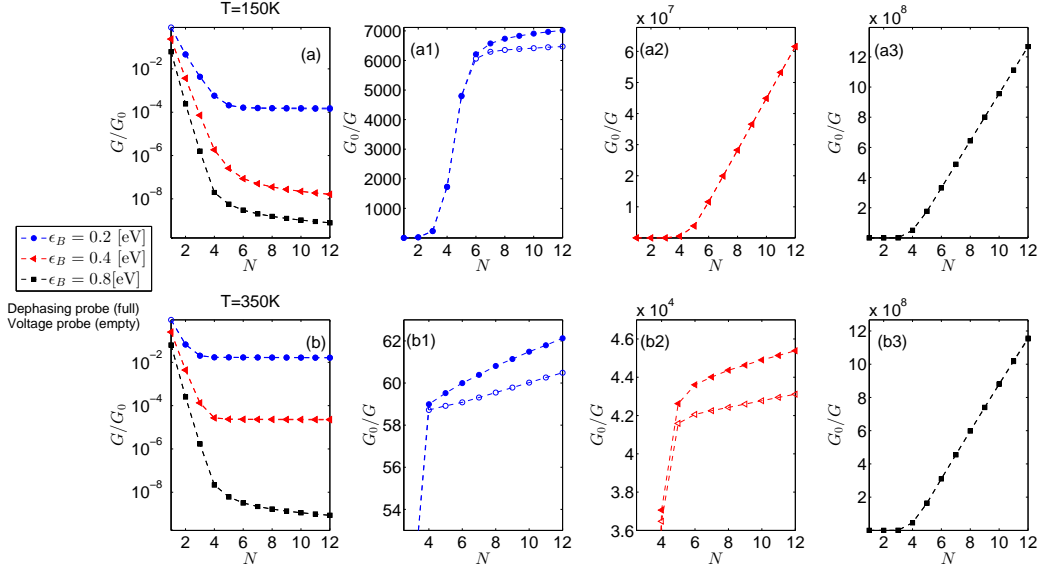}}}
\caption{
Analysis of bath-assisted conduction, $\gamma_d=1$ meV.
(a) At low temperatures $T=150$ K and for high bridges an activation-less transport takes place.
(b) At high temperatures $T=350$ K the conductance is enhanced by a thermal factor.
Panels (a1)-(a4) and (b1)-(b4) expose the ohmic-like behavior in long wires,
$v=0.05$, $\gamma_{L,R}=0.2$, $\Delta \mu=0.01$ eV,
dephasing probe condition (full symbols), voltage probe (empty symbols).
}
\label{FigTD}
\end{figure*}


\subsection{Junction's energetics}
\label{Results-E}

To further expose the dependence of the hopping conductance on the junction's energy parameters,
we study the behavior of $G$ with $\epsilon_B$ (set relative to the Fermi energy),
inter-site tunneling $v$, and the hybridization energy $\gamma_{L,R}$. As before, we restrict parameters
to the regime $\epsilon_B>v,\gamma_{L,R}$.

In Fig. \ref{Fig5} we display
the conductance as a function of $\epsilon_B$ for $N=8$ and $N=12$ under nonzero dephasing
and reveal a crossover from $G\sim e^{-\beta \epsilon_B}$ to $G\sim\frac{1}{\epsilon_B^4}$,
corresponding to the transition between a ballistic behavior (no length dependence)
dominating at small $\epsilon_B$, to off-resonance
dephasing-induced hopping conduction.

We now analyze more carefully the dependence of the conductance on temperature and bridge energy.
We recall that two limiting mechanisms prevail in the absence of dephasing effects:
deep tunneling conduction and ballistic transmission. Ballistic motion dominates in long wires, and it carries
the factor $\exp(-\beta \epsilon_B)$, see Eqs. (\ref{eq:Gr1})-(\ref{eq:Gr2}),
reflecting the thermal occupation of electrons in the metal at the bridge energy $\epsilon_B$.
Since it is non-dissipative, the ballistic component does not decay with distance, see Fig. \ref{Fig2}.
Dephasing-assisted conductance has a nontrivial temperature dependency due to the contribution of
many electrons in the metal.
In Fig. \ref{FigTD}(a)  we find that at low temperatures
the conductance  is missing an Arrhenius activation factor,
 $G(\epsilon_B=0.4)/G(\epsilon_B=0.8)\sim \left(\frac{0.8}{0.4}\right)^4=16$,
while a different behavior takes place at high temperatures $T=350$ K, as demonstrated in panel (b).
Here, in conjunction with an ohmic-like decay,  (focus on the red and black lines)
$G(\epsilon_B=0.4)/G(\epsilon_B=0.8)\sim 10^5$, reflecting an activated-thermally enhanced conductance.
We confirm in panels (b1)-(b3)
that an ohmic-like behavior develops with $1/G\propto N$.

We also compare in Fig. \ref{FigTD} the operation of the dephasing and voltage probes. We find that at this low bias limit
 ($\Delta \mu=10$ meV) transport characteristics are almost identical under either probes: the tunneling-to-hopping crossover
takes place at a similar molecular bridge size,
and the values for the conductances are close.
The two probes show some deviations only when the bridge height is taken relatively low
$\epsilon_B=0.2$ eV, with the dephasing probe yielding lower values for conductance (larger resistance per site)
in the hopping regime.

\subsection{Hopping conduction: construction of Eq. (\ref{eq:TH}).}
\label{Results-probe}

Based on the simulations reported above, as well as additional results included in the Appendix,
we suggest the hopping conductance, Eq. (\ref{eq:TH}),
%
$G_{H}\sim G_0
A(T) \frac{\gamma_d^2 v^2}{\epsilon_B^4}\frac{1}{ N +l}$,
%
with the dimensionless coefficient $l$, positive or negative, see Fig \ref{Fig5}.
This expression constitutes
central characteristics of bath-assisted conduction: Ohmic-like conduction for long chains, and
the enhancement of $G$ with dephasing strength (for weak dephasing) and
with barrier parameters as $\frac{v^2}{\epsilon^4}$.
Recall that Eq. (\ref{eq:TH}) was developed assuming parameters in the range
 $\gamma_d, v, \beta^{-1},\Delta \mu  \ll \epsilon_B$, $\gamma<\epsilon_B$,

We can justify this hopping form with some qualitative arguments:
Electrons scatter into the molecule through the resonance most strongly hybridized with the $L$ metal,
and leave from the orbital most tightly coupled to the $R$ lead, overall contributing the factor
$\frac{v^2}{\epsilon^4}$.
On the molecule, the probes absorb and emit electrons, dispersing them between molecular orbitals,
thus the conductance scales as $\gamma_d^2$,
describing probe-molecule-probe transitions.
The contribution of probe-mediated scattering processes linearly grows with the number of molecular sites (probes),
thus the hopping resistance grows as $N$.
The overall temperature dependence in the process is
rather weak; the hopping conductance integrates electrons thermally populating many levels in the lead,
off-resonant as well as in resonance with molecular orbitals.

Both dephasing and voltage probes disperse electrons between molecular orbitals.
However, the dephasing probe does so while maintaining the number of electrons
within each energy interval fixed.
The voltage probe can
absorb/provide energy from/to electrons, and it can thus modify the energy distribution of electrons in the system.
At low bias, $\Delta \mu<\gamma_{\nu}, \beta^{-1}$, the two probes produce very similar conductance characteristics.


\subsection{Non-Markovian environment}
\label{Results-nonmark}

We implement a structured environment by using energy-dependent dephasing energies with
the Debye-Drude form, Eq. (\ref{eq:SE}), see Fig. \ref{Fig8}.
As expected, we find that the conductance in the hopping regime is reduced due to the non-Markovianity of the bath,
while the tunneling and ballistic regimes are unaffected.
The effect is clearly observed at low temperatures,
but it is significantly mitigated at higher temperatures, when the dominant mechanism is ballistic conduction.

\begin{figure}[htbp]
\vspace{0mm} \hspace{0mm}
{\hbox{\epsfxsize=80mm \epsffile{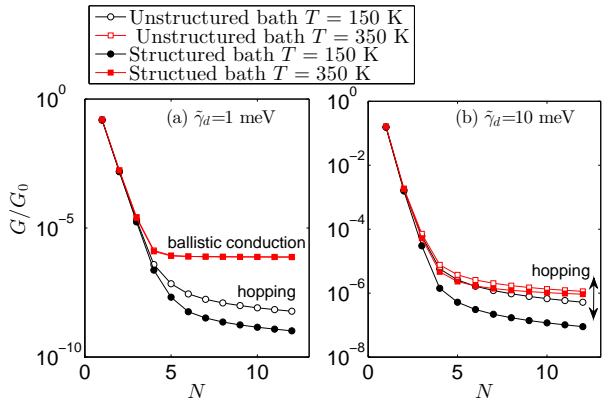}}}
\caption{
Electrical conductance as a function of length using structured [Debye-Drude form, Eq. (\ref{eq:SE})] and unstructured
($\gamma_d$ is a constant)
environments at high and low temperatures.
At low temperatures the conductance is sensitive to the cutoff frequency $\Omega=0.1$ eV.
We used $v=0.05$, $\epsilon_B=0.5$, $\gamma_{L,R}=0.2$, $\Delta\mu=0.01$ eV and
$\tilde \gamma_d=1, 10$ meV as indicated,
dephasing probe condition.
For unstructured environments we used $\gamma_d=1, 10$ meV, independent of energy.
}
\label{Fig8}
\end{figure}


\section{Beyond linear response: Current-voltage characteristics}
\label{Svol}

As a further application of our method, we present in Fig. \ref{Fig9} $I-V$ traces of molecular wires of size
$N=7$. These simulations were performed using the dephasing probe condition since the present
implementation of the voltage probe is applicable only in the linear response regime.
The voltage probe technique can be extended beyond that, by solving sets of nonlinear equations,
see e.g. Ref. \cite{SalilEPJ}.

Considering molecules under a large applied bias, we need to first determine the details of the
electrostatic potential profile across the junction.
As was discussed in several works \cite{profile-mujica, profile-datta,TVSPRL,Beebe},
in the weak metal-molecule coupling limit the applied voltage typically
drops only at the metal-molecule contacts (symmetrically or asymmetrically), thus
the barrier maintains its equilibrium form. 
In contrast, at intermediate-strong molecule-metal coupling the applied bias largely drops across the molecule.
Assuming a linear drop, the molecular barrier transforms from  a
rectangular to a trapezoidal shape, then to a triangular form when the bias exceeds the barrier height.
This transition from a ``direct tunneling" to ``field emission
transport" (Fowler-Nordheim tunneling) was explored experimentally
with transition voltage spectroscopy (TVS), when plotting $\log(I/V^2)$ as a function of $1/V$
to extract the barrier height from the minima of this plot.

Here, we assume a linear potential profile and shift the energies of sites $n=1,2,...,N$ accordingly,
\bea
\epsilon_n= \epsilon_B+ \frac{\Delta \mu}{2} - \frac{\Delta \mu (n-1)}{N-1}.
\label{eq:profile}
\eea
Using this model, $I-V$ characteristics at different values for the dephasing strength are plotted in
Fig. \ref{Fig9}, as well as examples for the potential profile at low, intermediate, and large biases,
see inset.
The $I-V$ curves reveal the followings: Up to approximately 0.1 V a linear response behavior is maintained.
The current is strongly enhanced when the voltage bias approaches a molecular resonance (located at $\sim 0.5$ eV).
Beyond that, $\Delta \mu\gtrsim 2\epsilon_B$, the current saturates.
The effect of the thermal bath on the $I-V$ characteristics is substantial.
At low bias, dephasing effects assist electron conduction,  $G\propto \gamma_d^2$.
Further, in resonant situations while the current is saturated for $\gamma_d=0$,
it displays a negative differential conductance under dephasing.
At large bias, the  effect of dephasing on current is non-monotonic and deserves further attention.

We point that a detailed TVS analysis, or the application of similar \cite{profile-markussen}
or alternative  \cite{profile-Cahen} measures,
are not particularly illuminating here since we had adjusted
the barrier, from a rectangular to a triangular shape, by hand, imposing the model (\ref{eq:profile}).
A TVS analysis or related investigations are interesting for e.g.,
exposing the evolution of the barrier height with size and contact properties in different
families of molecules, such as alkanethiols chains or aromatic thiols systems \cite{Beebe}.

\begin{figure}[htbp]
\vspace{0mm} \hspace{0mm}
{\hbox{\epsfxsize=80mm \epsffile{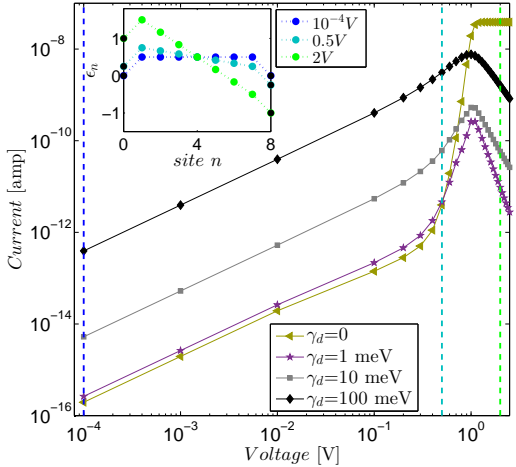}}}
\caption{
$I-V$ traces for a molecular bridge seven units long
with $T=298$ K, $v=0.05$, and $\gamma_{L,R}=0.2$ eV.
The bridge energies shift with the applied bias in a linear manner,
Eq. (\ref{eq:profile}), see examples in the inset. Sites $n=0$ and $n=8$
correspond to the chemical potentials at the contacts, $\mu_L$ and $\mu_R$, respectively.
}
\label{Fig9}
\end{figure}



\section{Comparison to experiments: ONI wires}
\label{Sexp}

In this Section we examine experimental results using the LBP method, to demonstrate the utility of the technique.
Ref. \cite{Exp-Frisbie-jacs} reported the resistance and current-voltage characteristics of
conjugated oligonaphthalenefluoreneimine
(ONI) molecular wires of different lengths, up to 10 nm with 10 repeating units, measured using conducting-probe atomic force (CP-AFM)
microscopy.
Based on the length-dependent and temperature-dependent resistance behavior, it was argued
that short wires conduct via a coherent-tunneling transport mechanism, while around 4 nm a transition in transport
behavior, from tunneling to thermally-activated hopping, took place.
Below, we use the data reported in Ref. \cite{Exp-Frisbie-jacs} and extract relevant parameters for our model.
We then demonstrate in Figs. \ref{FigE1}-\ref{FigE2} that by tweaking
the dephasing rate and the temperature we can reasonably reproduce experimental results of transport in ONI wires,
further providing some insights.

It should be emphasized that values for resistance reported in Ref. \cite{Exp-Frisbie-jacs} correspond
to a monolayer with $M\sim 100$;
$M$ is an estimate for the number of molecules under the measuring tip \cite{Exp-Frisbie-jacs}.
If we assume a linear scaling of the conductance, $G(M)\sim M G(1)$, the resistance should obey $R(M)\sim M^{-1} R(1)$.
Certainly, this is an approximation which is not always justified;
the conductance of a single-isolated wire $G(1)$ may significantly deviate from the monolayer conductance scaled-down, $G(M)/M$,
due to cooperative effects, inter-wire couplings and
substrate-mediated coupling. These effects may increase or decrease the conductance per wire
 \cite{Arik,Weitao,Reuter-lett,Reuter-nano11,Reuter-nano12}.
A careful study of such effects with the LBP method is of interest.
Here instead we use a naive-linear scaling approximation:
We take resistance values reported in Figs. 6 and 8 of Ref. \cite{Exp-Frisbie-jacs}
and multiply them by $M=100$; see the symbols ($+$) in Figs.
\ref{FigE1} and \ref{FigE2}.
Below we explain which parameters in our modelling are affected by this simplified scaling, and if so,
the impact on analysis.

To uncover model parameters from Ref. \cite{Exp-Frisbie-jacs}, $\epsilon_B$, $v$, $\gamma_{L,R}$, and $\gamma_d$, we
first determine the unit length of the ONI wires to be $a\sim 1$ nm.
We then obtain our parameters by following three steps:

(a) {\it Bridge energetics.} In the tunneling regime $N=1-3$,
 $G\sim e^{-\kappa Na}$, with  $\kappa =2.5 $ nm$^{-1}$ \cite{Exp-Frisbie-jacs}.
Using the relation $\kappa\equiv \frac{2}{a}\ln|\frac{\epsilon_B}{v}|$ [see Eq. (\ref{eq:exp})],
we resolve the ratio $\epsilon_B/v=3.4$.
To obtain the bridge height $\epsilon_B$ and inter-site energies, separately,
we note that the activation energy in ONI wires was approximated by $E_A\sim 0.55$ eV from an Arrhenius plot \cite{Exp-Frisbie-jacs}.
Furthermore, in Fig. \ref{Fig3} we found that while the activation energy linearly follows the bridge height,
the slope deviates from unity,
$\epsilon_B\sim 1.4 E_A$ for $\gamma_d\sim 5$ meV. We thus select the value $\epsilon_B=0.8$ eV,
and immediately receive $v=0.22$ eV. Note that these estimates are not affected by the scaling employed. Note also that the ballistic conductance is very sensitive to temperature near room temperature,
and therefore is also very sensitive to the bridge height/activation energy.
Even when using our ``corrected" $\epsilon_B=0.8$ value
for bridge energy, we still observe excess ballistic conductance near room temperature, and can thus only match experimental results
at a lower, ``effective temperature".

(b) {\it Hybridization energy.} We resolve $\gamma_{L,R}$ from Eqs. (\ref{eq:Gc})-(\ref{eq:exp})
(using $G_0=e^2/\pi\hbar$ to accommodate spin degree of freedom).
We assume that this contact energy is identical at the
two interfaces, and employ $\gamma_{\nu}=\sqrt{G/G_0 \times e^{\kappa N a}}$.
From the measured exponent $\kappa =2.5 $ nm$^{-1}$
we receive $\gamma_{\nu}\sim 0.018$ eV.
This estimate will vary if we adjust the scaling for $G(M)$, see discussion below.

(c) {\it Dephasing strength.}
We approach this parameter, characterizing electron-vibration interaction in the molecule,
by testing different values (uniform along the wire) in the range $\gamma_d=0-100$ meV. In
Fig. \ref{FigE1} we show
that very small dephasing values $\gamma_d=0-0.2$ meV
reasonably reproduce both the exponent in the tunneling regime and the slope in the hopping-ohmic domain,
assuming the junction's temperature is in the range 280-300 K.
As mentioned above, a very good agreement is reached by slightly reducing the temperature; alternatively, higher bridge heights could be tested.

With these molecular parameters, we turn now to the Arrhenius plot of ONI wires, as reported in Ref. \cite{Exp-Frisbie-jacs}.
Figure \ref{FigE2} depicts ($+$) the resistance as a function of inverse temperature for an $N=7$ ONI wire,
taken from Ref. \cite{Exp-Frisbie-jacs}, again linearly scaled down to the single molecule value.
We find that our results agree fairly well with experimental data under a coherent transport mechanism, and
we resolve the activation energy $E_A\sim 0.4$ eV from simulations, compared to the experimental value $E_A\sim 0.5$ eV.
We try to reproduce this data using the parameters resolved in the discussion above, but find out
that while at room temperatures and above, the experimental data is reasonably captured by our simulations with
$\gamma_d\sim 3-4$ meV,
at lower temperatures the experiment is reproduced with $\gamma_d=0$ meV. A possible explanation for this behavior
is that the dephasing rate in fact depends on temperature, $\gamma_d(T)$, and its value should be taken small
$\gamma_d\ll 1$ meV at low temperatures, $T=240$ K, while at room temperature it reaches values identified above.

Our analysis is based on the linear scaling assumption. It was demonstrated in Ref. \cite{Arik}
that this  holds only beyond a certain molecular-island size of a few tens of molecules.
Scaling the experimental results of Ref. \cite{Exp-Frisbie-jacs} by $M\neq 100$
would alter our estimate for $\gamma_{\nu}$ and $\gamma_d$; recall that
$G\propto \gamma_L\gamma_R$ in the tunneling regime and $G\propto \gamma_d^2$ in the hopping limit,  see Sec. \ref{Sres}.
As a limiting case, if reported values in Ref. \cite{Exp-Frisbie-jacs} were to correspond to a single molecule
 rather than to $M\sim 100$ molecules, our simulations would fit the data with the same values for the  bridge
energetics, $\epsilon_B=0.8$ eV, $v=0.22$, but with
(factor of 10 larger)  $\gamma_L=\gamma_R=0.18$ eV and $\gamma_d\sim$ 2 meV.
These values should serve as an upper estimate for dephasing strength.

Concluding our observations,
the probe technique under weak dephasing as a tweaking parameter
can capture the turnover behavior between superexchange and hopping regimes in molecular wires,
given the correct system parameters as inputs. We also see from Fig. \ref{FigE1}(a2),
as well as in  Fig. \ref{FigE2}, that in order to match the experiment across all temperatures and lengths,
the tuning parameter $\gamma_d$ must depend on temperature, and that one should take into account corrections
to the bridge height as it relates to the activation energy.


\begin{figure}[htbp]
{
\hbox{\epsfxsize=100mm \epsffile{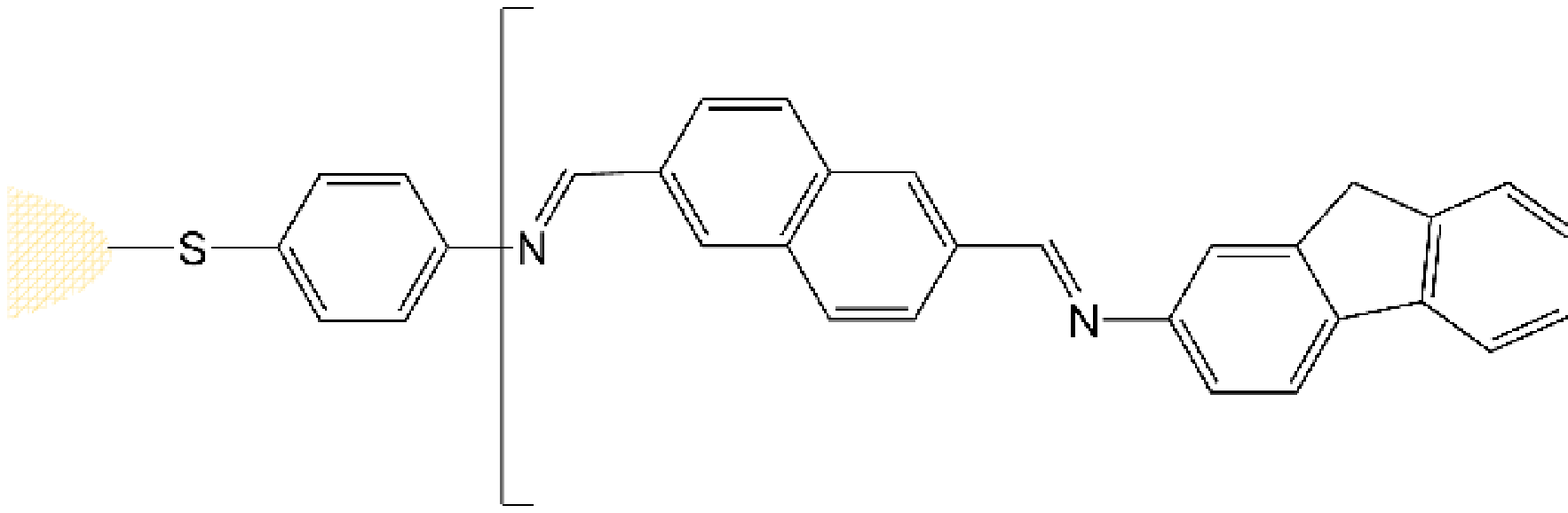}}
\vspace{-38mm}}
{\hbox{\epsfxsize=80mm \epsffile{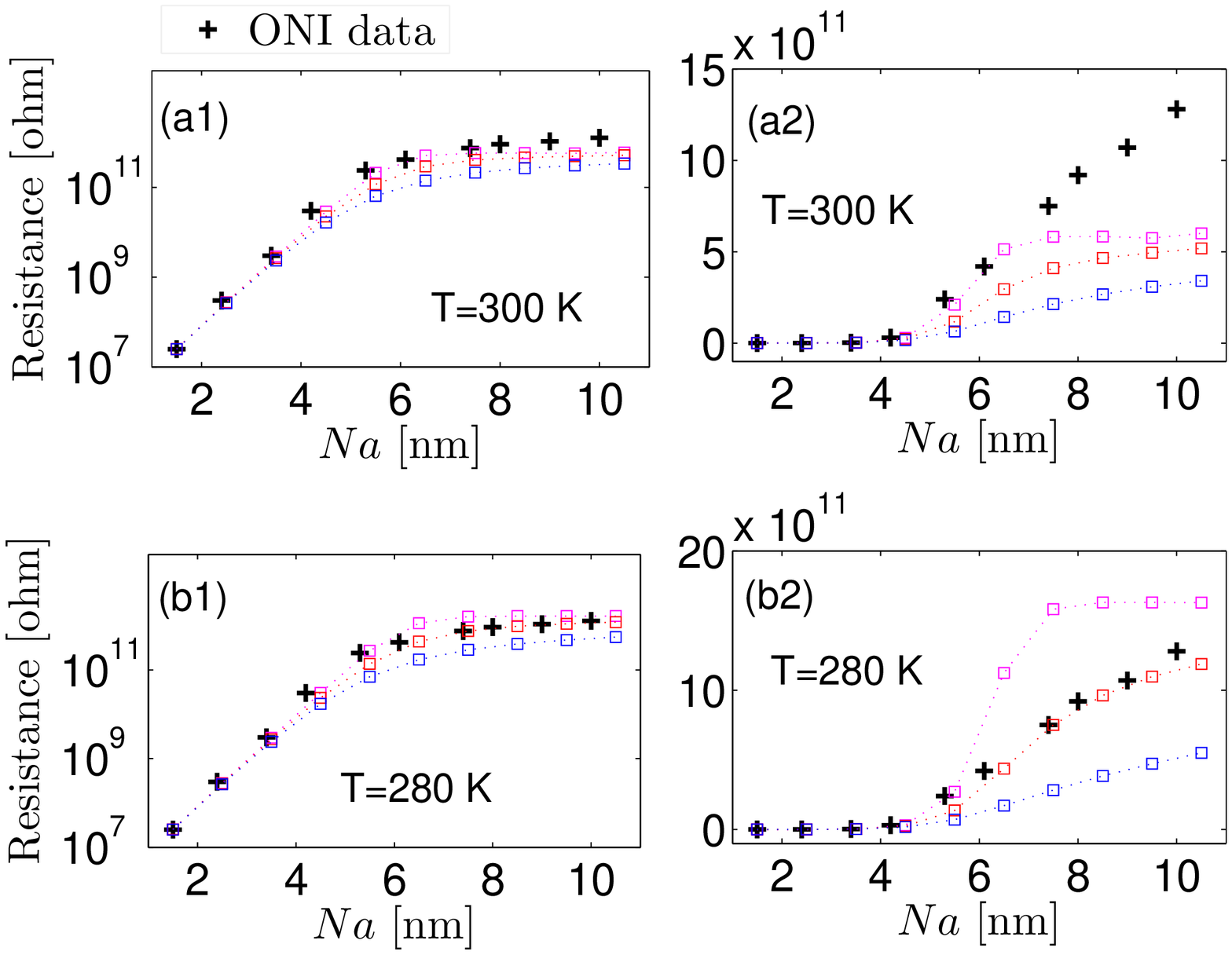}}}
\caption{
Analysis of experimental results. 
We extracted data for the resistance of an ONI monolayer with $M\sim 100$ from Fig. 6 of Ref. \cite{Exp-Frisbie-jacs},
and multiplied it by $M=100$ to approximate resistance per molecule.
Top: a single ONI wire (length denoted by $P$ here, rather than $N$, to eliminate confusion).
(a)-(b) 
Experimental data ($+$) \cite{Exp-Frisbie-jacs}
as a function of molecular length, compared to
probe-method simulations with $\gamma_d=0, 0.2, 0.5$ meV,
top to bottom (dotted lines with empty squares) assuming (a1)-(a2) $T=300$ K and
(b1)-(b2) $T=280$ K.
Panels (a1)-(b1) illustrate the tunneling region for $N=1-4$,
panels (a2)-(b2) uncover the  ohmic behavior for $N>5$.
Dephasing-probe simulations were performed with $\epsilon_B=0.8$ eV, $v=0.22$ eV, $\gamma_{L,R}$=0.018 eV,
applied voltage $\Delta \mu=0.01$ eV and dephasing strengths in the range $\gamma_d=0-1$ meV.
}
\label{FigE1}
\end{figure}

\begin{figure}[htbp]
\vspace{0mm} \hspace{0mm}
{\hbox{\epsfxsize=70mm \epsffile{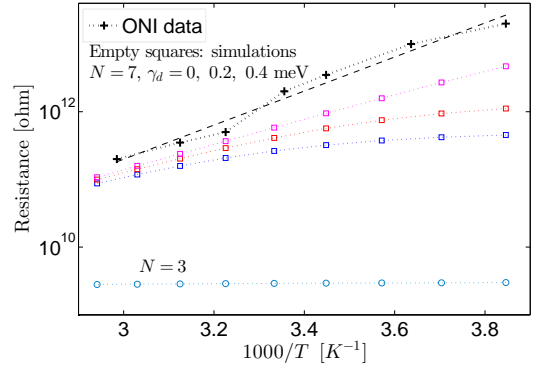}}}
\caption{Analysis of experimental results for the temperature dependence of conductance in ONI wires.
Experimental data ($+$) \cite{Exp-Frisbie-jacs} for $N=7$ is compared to
simulations with $\gamma_d$=0, 0.2, 0.4 meV (empty squares, top to bottom).
For reference, we also show the $N=3$ case ($\circ$) in which the resistance is
independent of dephasing in the range $\gamma_d\sim 0-5$ meV (consistent with the experiment).
Other parameters are the same as in Fig. \ref{FigE1}.
}
\label{FigE2}
\end{figure}

\section{Conclusions}
\label{Ssum}

We demonstrated that the Landauer-B\"uttiker probe technique can be used
to simulate electronic conduction in molecular junctions under the influence of a thermal environment.
The LBP approach can be incorporated with little effort into the commonly used Landauer-NEGF approach,
to phenomenologically introduce bath-assisted site-to-site hopping conduction.

Our simulations revealed different transport mechanisms: In the absence of dephasing effects we identified
deep-tunneling conduction in short wires and ballistic transport at high
temperatures, a result of resonant transmission.
Under finite dephasing strengths,
hopping conduction was the dominant transport mechanism in wires of $N>4$, showing temperature dependence which was generally weaker than the Arrhenius activation factor in the explored range of parameters.

Principal observations are:
(i) Length-dependence investigations of molecular conduction immediately pinpoint the tunneling-to ohmic turnover.
In contrast, activated conduction (or more generally, temperature dependent conduction)
may show up in both the coherent (ballistic) and incoherent-hopping regimes.
(ii) Kramers-like turnover behavior manifests itself under the LBP method.
Weak dephasing effects promote hopping conduction, but dephasing becomes detrimental to transport when $\gamma_d
\gtrsim\epsilon_B$.
(iii) Structured environments can be implemented within the LBP method, by using energy-dependent dephasing coefficients.
(iv) At large voltage biases, dephasing effects realize a negative differential conductance; this behavior
 is missing in the coherent limit.
(v) A closed-form expression for the hopping conduction was constructed, Eq. (\ref{eq:TH}),
within stated parameters.
(vi) The LBP method provided a semi-quantitative match for  length dependent resistance in ONI wires
reproducing the tunneling to hopping crossover, as well as the resistance's
temperature dependence- as long as we allow the dephasing strength to become temperature dependent.

B\"uttiker's probes offer an easily implemented mean to introduce dephasing and inelastic effects into molecular electronic
applications, yet, obviously they
do not absolutely correctly emulate physical process e.g., electron-vibration interaction.
While overall the LBP approach ensures the conservation of current
across the molecule, even within a small energy interval (dephasing probe),
one should remember that the probes are metal electrodes, collection of Fermi sea electrons
which themselves participate in the transport process.
Essentially, the quadratic dependence of the hopping conductance on the probe hybridization ($G_H\propto \gamma_d^2$)
reflects probe-molecule-probe electron scatterings,  
with an incoming electron from probe $n$ arriving at the molecule, leaving it to probe $n'$;
complementary processes ensure conservation of charge and energy, as necessary. In contrast, microscopic modelling of
electron-vibration interactions should yield to lowest order
a hopping conductance linear in the dephasing rate,
possibly modifying Eq. (\ref{eq:TH}), $\gamma_d^2\to \gamma_d/\beta$.
Thus, while the probes reproduce the hopping-ohmic nature of bath-assisted conduction,
the predicted dephasing and temperature dependence should be taken with caution.
To understand the correspondence of LBP results with physical-microscopic modelling,
one should compare our simulations to projection-operator approaches and first-principle techniques \cite{INFPIy}.

In future studies we will examine the suitability of the LBP technique
to describe the phenomenology of electronic function under environmental effects,
e.g. consider molecular thermoelectricity under thermal effects \cite{SegalthermoE,YelenathermoE}
and the  development of a diode behavior in symmetric molecules, induced by the surroundings \cite{Campos}.

\begin{acknowledgments}
This work was funded by the Natural Sciences and Engineering
Research Council of Canada and the Canada Research Chair Program.
We thank Professor Daniel C. Frisbie and 
Davood Taherinia for helpful discussions over experimental data, and for providing us
with their new results.
DS acknowledges Dr. Liang-Yan Hsu for interesting discussions over related Redfield calculations.
\end{acknowledgments}



\renewcommand{\theequation}{A\arabic{equation}}
\setcounter{equation}{0}  
\section*{Appendix A:  Supporting simulations for hopping conductance.}

In the main text we established Eq. (\ref{eq:TH}) by exploring
the dependence of the hopping conductance on  $\epsilon_B$, $\gamma_d$, temperature and length.
Here we complement this and examine the role of
inter-site tunneling $v$  and  metal-molecule hybridization on $G_H$.
In Fig. \ref{Fig6} we confirm that in the coherent-tunneling regime $G\propto v^{2N-2}$, but
at finite dephasing and for large $N$
we obtain $G\propto v^2$ for $v\ll \epsilon_B$, independent of molecular length (not shown).


\begin{figure}[htbp]
\vspace{0mm} \hspace{0mm}
{\hbox{\epsfxsize=70mm \epsffile{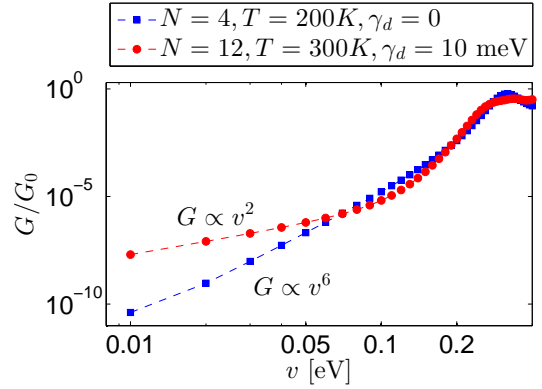}}}
\caption{
Electrical conductance vs. inter-site tunneling $v$. In the coherent tunneling limit
$G\propto \frac{1}{v^2}\left(\frac{v}{\epsilon_B}\right)^{2N}$.
In the hopping regime $G\propto v^2$. 
Parameters used are $\epsilon_B=0.5$, $v=0.05$, $\gamma_{\nu}=0.2$, $\Delta \mu=0.01$ eV.
}
\label{Fig6}
\end{figure}

\begin{figure}[htbp]
\vspace{0mm} \hspace{0mm}
{\hbox{\epsfxsize=70mm \epsffile{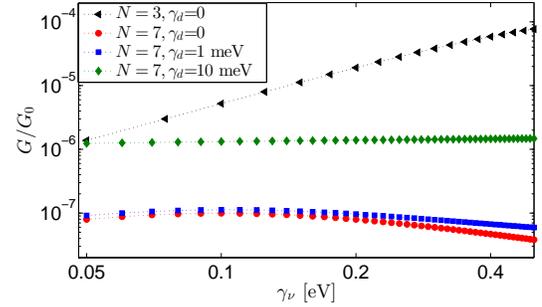}}}
\caption{
Electrical conductance as a function of the hybridization energy, $\gamma_L=\gamma_R$.
When $\gamma_d=0$  we resolve tunneling conductance for $N=3$, $G\propto \gamma_L\gamma_R$ ($\triangle$)
and ballistic motion for $N=7$ when $\gamma_{\nu}\ll \epsilon_B$ ($\circ$).
The conductance  dependency  on $\gamma_{\nu}$ diminishes upon increasing dephasing strength.
We used $T=300$ K, $\epsilon_B=0.5$, $v=0.05$, $\Delta \mu=0.01$ eV, dephasing probe simulations.
}
\label{Fig7}
\end{figure}


The hybridization energy $\gamma_{\nu}$ is expected to only mildly influence hopping conduction,
as the resistance should be determined
by the wire itself - electron scattering between probes- rather than by the interface of the molecule with the contacts.
This behavior is confirmed in Fig. \ref{Fig7}: Tunneling conductance ($N=3$, $\gamma_d=0$)
grows as $\gamma_{L}\gamma_R$,
see Eq. (\ref{eq:exp}), but in long chains with $\gamma_d\neq0$ the effect of the hybridization energy on $G$
is insignificant.

{}

\end{document}